\documentclass[sigconf]{acmart}

\AtBeginDocument{%
  \providecommand\BibTeX{{%
    \normalfont B\kern-0.5em{\scshape i\kern-0.25em b}\kern-0.8em\TeX}}}

\setcopyright{acmcopyright}
\copyrightyear{2018}
\acmYear{2018}
\acmDOI{10.1145/1122445.1122456}

\acmConference[Woodstock '18]{Woodstock '18: ACM Symposium on Neural
  Gaze Detection}{June 03--05, 2018}{Woodstock, NY}
\acmBooktitle{Woodstock '18: ACM Symposium on Neural Gaze Detection,
  June 03--05, 2018, Woodstock, NY}
\acmPrice{15.00}
\acmISBN{978-1-4503-XXXX-X/18/06}

\begin{document}
\title{Online Trajectory Prediction for Metropolitan Scale Mobility Digital Twin}

\author{Zipei Fan$^1$, Xiaojie Yang$^1$, Wei Yuan$^1$, Renhe Jiang$^1$, Quanjun Chen$^1$, Xuan Song$^{1,2}$, and Ryosuke Shibasaki$^1$}

\affiliation{%
  \institution{$^1$Center for Spatial Information Science, University of Tokyo}
  \country{Japan}
}
\affiliation{%
  \institution{$^2$SUSTech-UTokyo Joint Research Center on Super Smart City, Southern University of Science and Technology}
  \country{China}
}

\renewcommand{\shortauthors}{Zipei Fan et al.}

\begin{abstract}
Knowing "what is happening" and "what will happen" of the mobility in a city is the building block of a data-driven smart city system. In recent years, mobility digital twin that makes a virtual replication of human mobility and predicting or simulating the fine-grained movements of the subjects in a virtual space at a metropolitan scale in near real-time has shown its great potential in modern urban intelligent systems. However, few studies have provided practical solutions. The main difficulties are four-folds. 1) The daily variation of human mobility is hard to model and predict; 2) the transportation network enforces a complex constraints on human mobility; 3) generating a rational fine-grained human trajectory is challenging for existing machine learning models; and 4) making a fine-grained prediction incurs high computational costs, which is challenging for an online system. Bearing these difficulties in mind, in this paper we propose a two-stage human mobility predictor that stratifies the coarse and fine-grained level predictions. In the first stage, to encode the daily variation of human mobility at a metropolitan level, we automatically extract citywide mobility trends as crowd contexts and predict long-term and long-distance movements at a coarse level. In the second stage, the coarse predictions are resolved to a fine-grained level via a probabilistic trajectory retrieval method, which offloads most of the heavy computations to the offline phase. We tested our method using a real-world mobile phone GPS dataset in the Kanto area in Japan, and achieved good prediction accuracy and a time efficiency of about 2 min in predicting future 1h movements of about 220K mobile phone users on a single machine to support more higher-level analysis of mobility prediction.
\end{abstract}



\maketitle
\begin{figure}[th!]
\includegraphics[width=\linewidth]{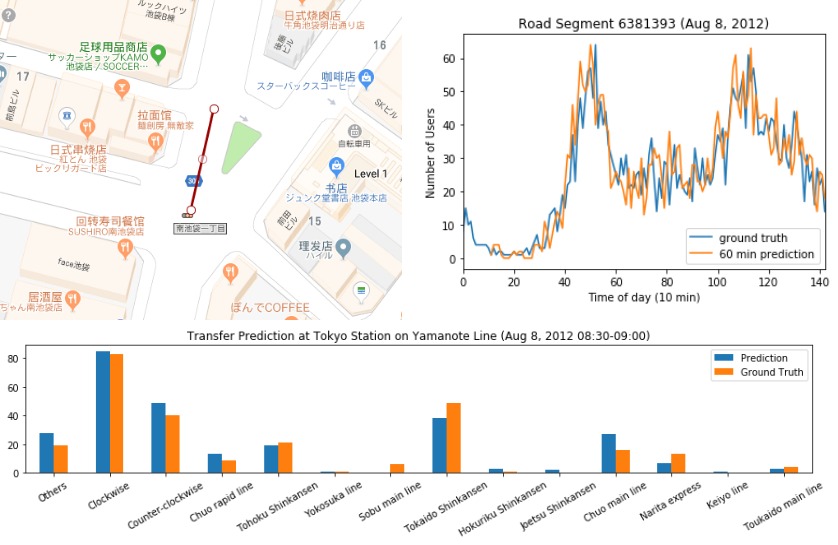}
\caption{Traffic volume analysis for a particular road segment (upper), and transfer prediction at Tokyo station on the Yamanote line (lower) in the mobility digital twin.}
\label{fig:fine_grained_prediction}
\end{figure}
\begin{figure*}
\centering
\includegraphics[width=0.96\textwidth]{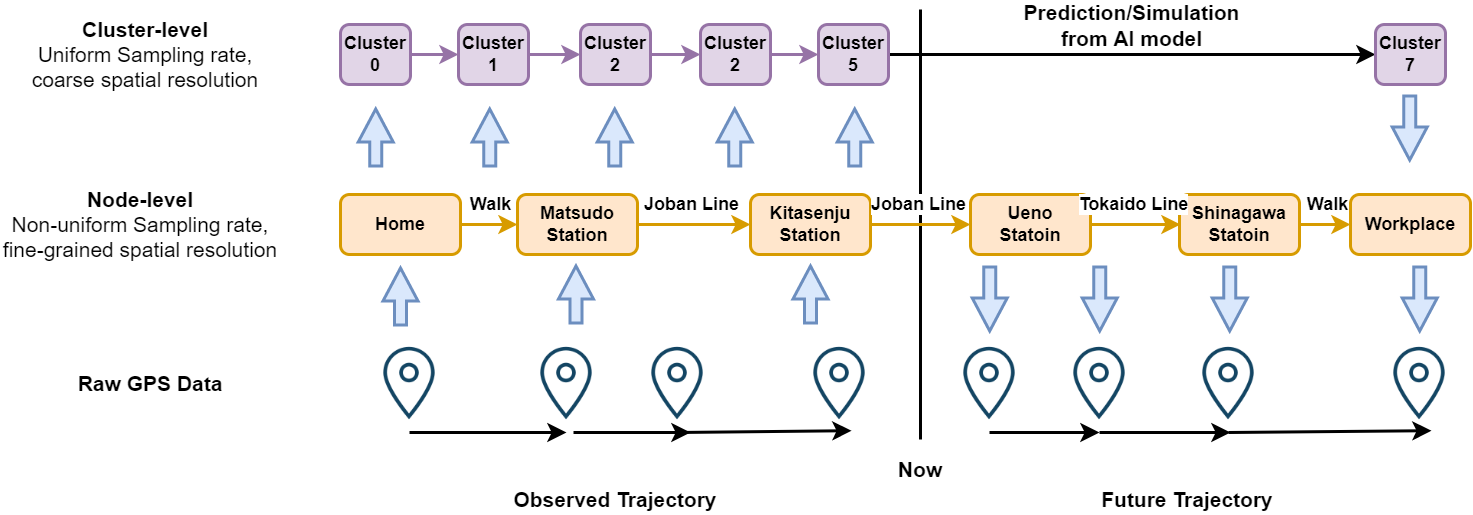}
\caption{Two staged mobility prediction framework.}
\label{fig:twostage}
\end{figure*}
\begin{figure}
\centering
\includegraphics[width=\linewidth]{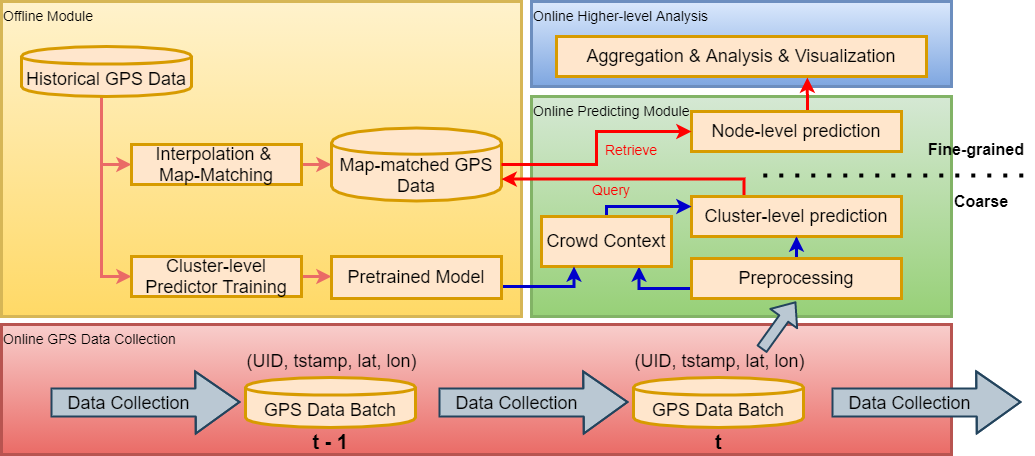}
\caption{Pipeline of our approach. Our approach comprises four modules: online GPS data collection module (red), online prediction module (green), offline module (yellow) and user interfaces for higher-level analysis module (blue).}
\label{fig:overview}
\end{figure}
\section{Introduction}
For crowd surveillance or traffic regulation systems, an accurate fine-grained prediction of human movements can help people make informed decisions and governments take timely countermeasures. Most existing studies predict human mobility at a coarse level, predicting either just the number of aggregated population or traffic volume or only the next-move/destination represented by the grid cell or coordinate, rather than a complete trajectory that is matched to the transportation network. Such coarse predictions are insufficient to support higher-level transportation predictive analysis that transportation bureaus or companies' are concerned with. Some examples of higher-level analysis are listed as:
\begin{itemize}
    \item What will the traffic volume on a particular road/railway segment be in 1h?
    \item How many people will get on/off and transfer in a specific station using a particular line (e.g., Tokyo station and Yamanote line)?
    \item Find out the potential sources/sinks of a gathering or dispersing pattern from the massive predicted human trajectories.
    \item How the users will move alternatively if the line stopped.
\end{itemize}
To this end, instead of modeling the mobility in aggregation, we need a more comprehensive virtual replica of the mobility in the physical world and make an accurate and efficient prediction of the future movements of the agents in the virtual space under the current urban status or simulated conditions. Recent progresses in digital twin, especially mobility digital twin \cite{deng2021systematic, wang2022mobility}, have shed light on a potential solution to this problem that senses, predicts, simulates and visualizes the mobility of a city in real-time. However, few studies have proposed practical solutions of making a fine-grained prediction of the massive agents in the virtual space accurately and efficiently. In this study, by collecting the raw GPS stream that makes a more lossless replication for the mobility of the city, we aim to make more informative predictions by refining our predictions from aggregated population/traffic to \textbf{individual}, from a grid-cell or coordinates represented by trajectory to \textbf{transportation network level}, and from predicting the next move or destination to a \textbf{complete future trajectory}. As shown in Fig. \ref{fig:fine_grained_prediction}, our fine-grained prediction can support further high-level analysis such as aggregation at the road network level to predict 1h ahead future traffic volume on a specific road segment, or aggregation at a station level to predict the future transfer behavior after people arrive at an interchange like the Tokyo station. The main challenges come from four aspects:

1) The predictor should be capable of capturing the differences of human mobility from day to day and adjust itself by perceiving the current crowd context. For example, a user goes to drinking places with a higher probability if we observe a higher population going to drinking places. Thus, we need to model both the sequential pattern of user trajectories and how the sequential pattern varies considering other trajectories within the crowd context.

2) The multimodal transportation network, especially in the urban areas, enforces complex constraints on human mobility. Considering road networks as an example, the road network in the Kanto area includes 2,902,380 road segments of 9 road types with different properties (access type, tolled or free, speed limits, etc.) and 2,035,726 road nodes. Existing prediction algorithms can hardly address so many complex constraints on human mobility in a feasible manner.

3) Learning to generate a rational human trajectory, which is a long sequence ($>$100) with a large vocabulary ($>$1K), is regarded as a challenging task in the machine learning community. Generating a 1h predicted trajectory at the transportation network level is an even more daunting problem because of the longer sequence length (as much as hundreds of road segments in 1h by car) and larger vocabulary (millions of road nodes or segments) size.

4) Inferring fine-grained trajectories from raw GPS trajectories requires heavy computation, especially to disambiguate the localization uncertainty and search for a reasonable route considering trajectory uncertainty, multimodality and different road properties.

To address these difficulties, we propose a two-stage human mobility predictor that stratifies coarse and fine-grained level movements, as shown in Fig. \ref{fig:twostage}. In the first stage, we encode the current global state of the city mobility as crowd context automatically, and predict the cluster-represented destination distribution at a coarse level. Note that, in this stage, we focus on addressing large-scale, long-term, and inter-user dependencies, while not considering local transportation network structure. This is because our local transportation choice is largely determined by long-term destination.

In the second stage, we develop a retrieval-based method to generate our predicted trajectories via a weighted sampling from the fine-grained historical trajectory database, where the weight is determined by the coarse level prediction in the first stage. Thus, we offload heavy computation to an offline phase; moreover our system is efficient and may practically generate timely, fine-grained future trajectories. The entire pipeline is shown in Fig. \ref{fig:overview}

We summarize our contributions as follows:
\begin{itemize}
    \item To make efficient prediction of the virtual replica of the mobility in physical world rather than aggregated data (e.g., traffic volume, population density), We propose a novel two-stage human mobility prediction framework that predicts metropolitan-scale human mobility at a fine-grained level in near real-time.
    \item To better encode the urban status, we propose a novel predictive model that perceives the citywide mobility as crowd context and adjusts the predictor in a meta-learning paradigm.
    \item We propose a novel probabilistic retrieval-based prediction approach that bridges cluster-level prediction with fine-grained future trajectory generation for the prediction or simulation in the mobility digital twin.
\end{itemize}
\section{Preliminaries}

In this section, we define the terms and concepts frequently used throughout this paper.

\begin{definition}[Raw GPS data] The raw GPS data can be formally described as follows: 
\begin{equation}
X=\left\lbrace \left( user\_id,\,time,\,latitude,\,longitude \right)\right\rbrace    
\end{equation}

Thus, the trajectory of each user $Tr^{u}$ can be defined as
\begin{equation}
Tr^{u} = x^u_{0}, x^u_{1}, \,\dots\, \quad x^u_{i} \in X
\end{equation}
where $x^u_{i}$ is the $i$-th record (sorted by time) of user $u$. 

To distinguish the online data stream and offline historical data, we use symbol prime \^ to denote those variables in or related to the historical trajectory database (e.g., the historical raw GPS data $\hat{X}$ and user ID $\hat{u}$ in the historical database).
\end{definition}

\begin{definition}[Cluster-level trajectory]\label{def:mapmatching}
We represent each trajectory on day $d$ as a cluster-level trajectory by dividing the continuous time and location representation into time slots $t$ and location cluster ID $C^u_{t}$.
\begin{equation}
    Trc^{u}_{d} = \left[\, C^{u}_{d, \,0},\, C^u_{d, \,1},\, \dots,\, C^u_{d, \,T - 1}\, \right]
\end{equation}
where $T$ is the number of time slices for one day. Location is represented by the index of cluster, and cluster-level trajectories are obtained in an online manner. In addition, we denote the collection of the cluster-level trajectories for all users as $TRC_{d} = \lbrace Trc^u_d \,\big|\, u \in U \rbrace$, where $U$ is the set of user IDs.
\end{definition}

\begin{definition}[Node-level Trajectory]\label{def:cluster_traj}
We incorporate the information obtained from a digital map to match the raw GPS trajectory with the transportation network. Thus, we get the node-level trajectory as
\begin{equation}
    Trn^{u} = (t_0, \, node_0), \, link_{0 -> 1}, \, (t_1, \, node_1), \, link_{1 -> 2}, \, \dots
\end{equation}
where $time_{i}$ and $node_{i}$ are the $i$-th time stamp and transportation node ID of the node-level trajectory, respectively. $link_{i - 1, \,i}$ is the link between $node_{i - 1}$ and $node_{i}$, which usually represents a segment of road or railway. 
\end{definition}
\begin{definition}[Crowd Context]\label{def:crowd_context}
We define the crowd context $\Phi_{t}$ at time $t$ to simplify the interdependence of the user trajectories $TRC[t - \Delta T: \, t]$ at time $t$. The crowd context encodes the current collective mobility patterns into a vector, and configures the cluster-level predictor to be more adaptive to the current mobility trend.
\end{definition}

\section{Our Approach}

\subsection{Cluster-level predictor}

\subsubsection{Cluster-level trajectory pre-processing}\label{sec:cluster_traj}
\begin{figure}
\centering
\includegraphics[width=\linewidth]{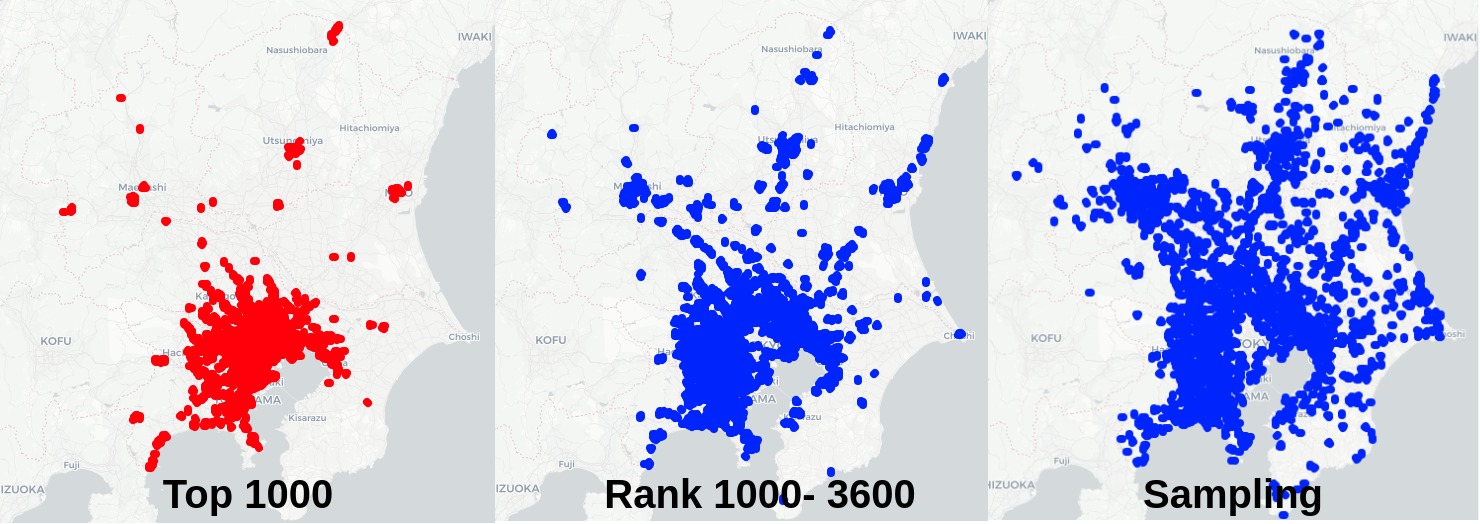}
\caption{Most frequently visited 1000 hexagons (left), top 1001-3600 hexagons (middle) and sampled 2600 hexagons (right) regarding visit frequencies.}
\label{fig:cluser_level_traj}
\end{figure}
To model long-distance and long-term dependencies from user trajectories, we need to transform raw GPS trajectory to cluster-level trajectory, as shown in Fig. \ref{fig:cluser_level_traj}. We obtain a cluster-level trajectory from raw GPS log data using a four-step process:

\begin{figure}[!t]
\centering
\includegraphics[width=\linewidth]{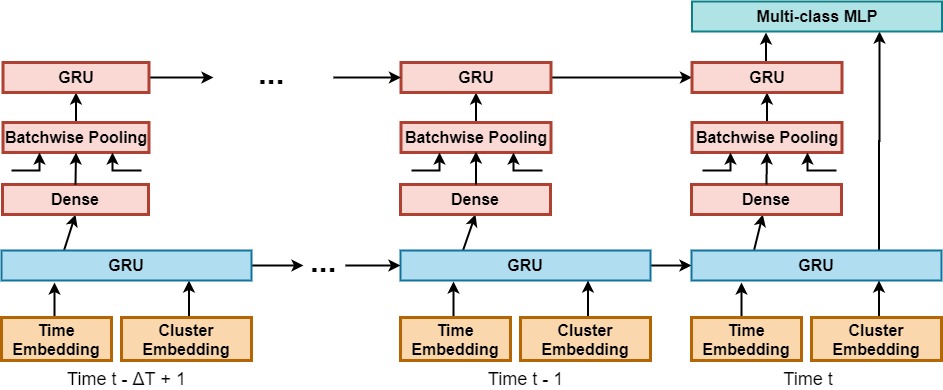}
\caption{Our proposed collective cluster-level mobility predictor. Red part shows how crowd context is used to enhance our prediction.}
\label{fig:cluser_predictor_network}
\end{figure}
\begin{itemize}
    \item Forwarding-fill the trajectory to obtain a uniform-sampling trajectory. Note that forward-filling is suitable for an online system because no future information is required.
    \item Index the trajectory location using Uber's Hexagonal Hierarchical Spatial Index (H3 index)\footnote{https://uber.github.io/h3/} with the 8th resolution (average hexagon area $0.74 km^2$).
    \item Count the visit frequency of the hexagons in the training data, and create two sets of hexagons: i) top 1000 most frequently visited hexagons, as shown in Fig. \ref{fig:cluser_level_traj} (left); and ii) sampled 2600 hexagons from the remaining hexagons with the probability proportional to their visit frequencies, as shown in Fig. \ref{fig:cluser_level_traj} (right).
    \item Simplify the H3-indexed trajectory by approximating the H3-indices that do not belong to Type i) using the nearest Type ii) hexagon. Thus, we use 3600 indices to represent all the locations in the dataset.
\end{itemize}

Therefore, we regularize the raw GPS trajectory from both spatial and temporal aspects. From the temporal aspect, we re-sample the non-uniform sampling trajectory to obtain a uniform sampling trajectory, which simplifies spatiotemporal features and accelerates the processing via batching. From the spatial aspect, we prefer discrete location representation, rather than continuous values (coordinates), because complex transportation networks make spatial dependency highly non-linear, especially if locations along the highways or railways can be regarded as singularities and can hardly be modeled by continuous functions.

In this study, we conduct a simplification of the H3-index with respect to the visit frequency. Many previous studies represent location by grid cell only. However, to model human mobility at a metropolitan scale, we need a more flexible resolution. Note that we apply a sampling strategy, rather than directly using the top 3600 hexagons for a compromise of spatial resolution. Compared with the geographical distribution of top 1001-3600 hexagons (middle) and sampled hexagons (right), we find that the top 1001-3600 is much more spatially concentrated; thus, the spatial resolution is relatively low compared with sampled hexagons concerning the fourth process step listed above.

\subsubsection{Cluster-level predictor}\label{sec:pretrain}

A basic model for cluster-level predictor can be described as
\begin{equation}
    p \left(Trc[t + \Delta T] \, \bigg| \,Trc[t - \Delta T\,:\, t]\right) = F\left(Trc[t - \Delta T\,:\, t]\right)
    \label{eq:base_prediction}
\end{equation}
where $F$ is the predictive function that considers the most recent $\Delta T$ records and predicts the probability of $\Delta T$-ahead future movement $Trc[t + \Delta T]$. A typical choice of $F$ is to utilize a gated recurrent unit (GRU) to model the sequential pattern of the most recent trajectory and a $SoftMax$ layer to transform the prediction into a probability distribution. This can be described as
\begin{equation}\label{eq:base_gru}
    \begin{array}{cl}
h_0 =& \textbf{0}\\
h_{\tau}=& GRU \left( \left[ EL\left(Trc[t - \Delta T + \tau]\right), ET\left( t \right) \right],\, h_{\tau - 1} \right)\\
o_T=& SoftMax\left( MLP\left(h_{\Delta T})\right) \right)
    \end{array}
\end{equation}
where $\tau = 1, \dots, \Delta T$, $h$ is the hidden state of the GRU at each time step, and $o$ the output vector representing the distribution of $Trc_{t + \Delta T}$. We use an embedding layer $EL$ with the vocabulary size of the number of clusters to map the cluster ID to an $N_{EL}$-dimensional vector. Similarly, the time-of-day is mapped to a $N_{ET}$ using the embedding layer $ET$.

\begin{figure}[!t]
\centering
\includegraphics[width=0.95\linewidth]{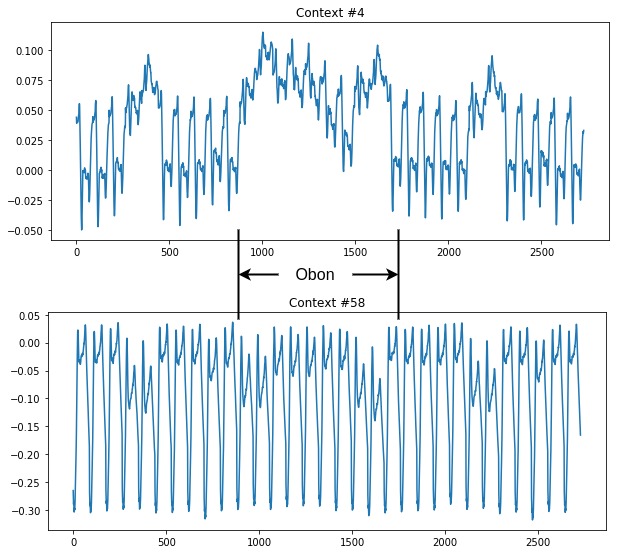}
\caption{Visualization of the 4th and 58th dimensions of crowd context changing with time}
\label{fig:context_dimension}
\end{figure}

However, the basic model does not consider the interdependence of the user trajectories. For example, when we observe some trajectories gathering at the central office area, it will increase our belief that it is a regular weekday, and that other users will also have a higher probability of going to their workplace. Consequently, we add flexibility to the predictive function $F$ that can adjust itself according to the current crowd context $\Phi_{t}$, characterizing the spatiotemporal features from all contemporary user trajectories $TRC[t - \Delta T:\, t] = \left\lbrace Trc^{u}[t - \Delta T\,:\, t] \,\bigg|\, u \in U\right\rbrace$. Thus, following a meta-learning paradigm, Equation \ref{eq:base_prediction} can be rewritten as
\begin{equation}
    p \left(Trc[t + \Delta T] \, \bigg| \,Trc[t - \Delta T\,:\, t], \,\Phi_{t}\right) = 
    \mathcal{F}\bigg( \Phi_{t} \bigg) \bigg(Trc[t - \Delta T\,:\, t]\bigg)
\end{equation}
where $\mathcal{F}$ utilizes the crowd context $\Phi_{t}$ to characterize the current crowd trend, and generates the adaptive predictor $\mathcal{F}\bigg( \Phi_{t} \bigg)$, which is equivalent to a flexible $F$ in Equation \ref{eq:base_prediction}. That is, we use the adaptive predictor $\mathcal{F}\bigg( \Phi_{t} \bigg)$ that changes with the crowd context $\Phi_{t}$ to replace the fixed predictor $F$ in our proposed method. Thus, our proposed predictor can exploit the periodic patterns we learned from the training data, can learn how to cope with the fluctuations in human mobility from the training data, and thus, is more robust to irregular human mobility.

Note that the crowd context $\Phi_{t}$ is calculated from $TRC[t - \Delta T:\, t]$, which is an unordered set of instances (trajectories); therefore, we need to choose a permutation-invariant function (pooling function) for estimating the crowd context by aggregating over all the instances. $Mean$ and $Max$ are the two most widely used and time-efficient permutation-invariant functions. $Max$ calculates the maximum value of each dimension of the feature vectors, and $Mean$ averages all these feature vectors. We will show in our later experiments (Fig. \ref{fig:cross_entropy_evaluation}) that $Mean$ pooling function outperforms $Max$. This is because $Mean$ is more capable of reflecting the crowd proportion, while $Max$ is better at testing the existence of the set. Let us assume that a feature describes whether a user is heading for his/her workplace. $Mean$ pooling can easily distinguish between 1\% and 20\% population, while $Max$ pooling can only test whether there is someone who is headed to work.

Note that, from Equation \ref{eq:base_gru}, the crowd context should be computed from the entire user set. However, it is ineffective to conduct mean pooling over the entire user set literally. In practice, we increase the size of the mini-batch (we set it to 4096 in our experiments) in our training phase to use the average of the mini-batch to approximate the crowd context; meanwhile, in the testing phase, we pre-compute and apply mean pooling to the crowd context over all the users.

To encode the sequential pattern of crowd context, we use another GRU network (red part in Fig. \ref{fig:cluser_predictor_network}) and merge with the hidden state from the GRU (blue part in Fig. \ref{fig:cluser_predictor_network}) defined in Equation \ref{eq:base_prediction}. The network structure of our proposed cluster-level predictor is shown in Fig. \ref{fig:cluser_predictor_network}.

Fig. \ref{fig:context_dimension} shows how the crowd context encodes the periodic and irregular facets of human mobility. We select two dimensions (4th and 58th) from the crowd context, and observe how the value of these two dimensions change during the testing phase. In the upper figure, the 4th dimension of the crowd context is more sensitive to irregular human mobility, especially as it clearly distinguishes the Obon festival week from regular weekdays. By contrast, the 58th dimension of the crowd context (lower) is less sensitive to irregular aspects of human mobility, while characterizing more periodic mobility patterns, even demonstrating some differences between the Obon festival week and a regular week. Therefore, every dimension of our crowd context encodes a different facet of large-scale human mobility, which is complementary with each other and supports our cluster-level predictors to simultaneously cope with both periodic and unexpected patterns underlying human mobility.

\subsection{Fine-grained historical trajectory database} \label{sec:interpolation}
Trajectory interpolation and map matching is the key process bridging GPS coordinates and transportation nodes and links. However, this procedure is very time-consuming and relies heavily on future information to disambiguate the uncertainties, as shown in Fig. \ref{fig:map_interpolation}. We can hardly determine whether the user is on the railway (blue) or road (black) until we obtain the information of his/her next move (blue cross). Thus, it is difficult to interpolate and map-match the raw GPS trajectory in real-time for an online system.

Therefore, we propose a retrieval-based prediction approach that offloads the heavy computation task to the offline phase. In the rest of this subsection, we will describe how to acquire a node-level trajectory and how to build a fine-grained historical trajectory database for online querying.

\subsubsection{Node-level trajectory}

\begin{figure}[!t]
\centering
\includegraphics[width=\linewidth]{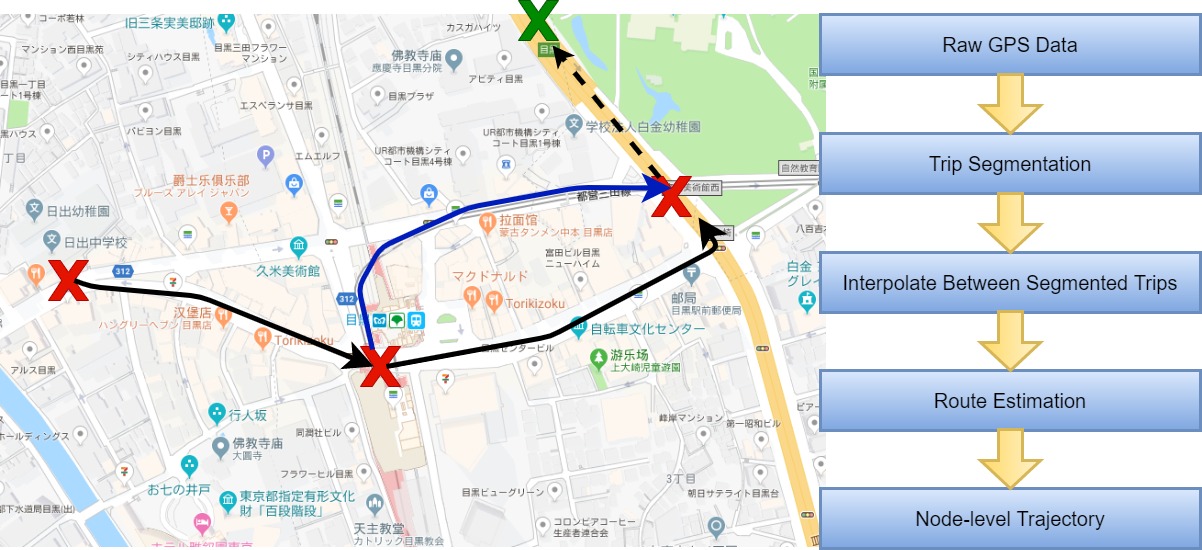}
\caption{Illustration of using future information to disambiguate localization uncertainty (left); the key steps of trajectory interpolation and map-matching (right).}
\label{fig:map_interpolation}
\end{figure}

Based on the methods introduced in \cite{Zheng:2008:UMB:1409635.1409677}, we conduct a three-step process to get a node-level trajectory as shown in Fig. \ref{fig:map_interpolation}:

1) Trip segmentation: stay points (with a maximum moving distance of 300m and minimum stay period of 15 min) are extracted from time-series GPS data based on spatiotemporal features, and moving state (MOVE or STAY). The transportation mode (WALK, BICYCLE, CAR, TRAIN, OTHER) is determined and partitioned for each trip based on the methods in \cite{Zheng:2008:UMB:1409635.1409677}.

2) Interpolation between segmented trips: owing to the sparse characteristics of raw GPS trajectories, there are blank gaps between segmented trips. Considering the moving distance (threshold = 200m), gap length (threshold = 1h), and the stay/move states at the two ends of the gap, we fill the blank following the rules.
\begin{itemize}
    \item If a long moving distance is presented, we insert a move state in the mid-term.
    \item If a state transition is presented, we determine the exact transition time estimated from the trajectories at the moving state.
    \item We merge consecutive STAY->STAY trips or MOVE->MOVE trips with the same transportation modes.
\end{itemize}

3) Route estimation: in order to snap the GPS points to transportation network nodes and estimate a more accurate travel time, we conduct a route search for each trip with the MOVE state depending on the transportation mode and road/railway type (highway, national road, railway of local/express train, etc.).

This above procedure is difficult for an online system because future information is critical to disambiguate the uncertainty, especially for sparse GPS data with low and non-uniform sampling rates and positioning errors. Moreover, the route estimation step is a non-deterministic problem and thus requires a large computational resources. Thus, we aim at obtaining good quality fine-grained historical trajectories in the offline module. Furthermore, in the following parts of this paper, we will show how it can be used in an online system in our retrieval-based prediction approach.

\subsubsection{Creating trajectory database}
By obtaining the fine-grained node-level trajectories, we create the key-value store historical trajectory database as:

\begingroup
\small
\begin{equation}\label{eq:database_definition}
    DB\left[{C}\right] = \left\lbrace\left( \hat{u}, \,\hat{d},\, \hat{t} \right)\, : \, Trn^{\hat{u}}_{\hat{d}} \left[\hat{t}: \hat{t} + \Delta T \, \right] \:\bigg|\: Trc^{\hat{u}}_{\hat{d}}[\hat{t} + \Delta T] = C\right\rbrace
\end{equation}
\endgroup

where node-level trajectory $Trn^{\hat{u}}_{\hat{d}}$ of user $\hat{u}$ on day $\hat{d}$ is partitioned into slices with time interval $\Delta T$. Each slice has key $\left( \hat{u}, \,\hat{d},\, \hat{t} \right)$. The database is partitioned by the destination cluster $Trc^{\hat{u}}_{\hat{d}}[\hat{t} + \Delta T]$, corresponding to our cluster-level prediction.

For an efficient query of historical trajectories, we map the historical trajectory to a Euclidean metric space considering two principles: 1) trajectory spatiotemporal continuity, meaning that the predicted trajectory should start close to where the observed trajectory ends; and 2) periodicity, meaning that the trajectories at the same time-of-day should share similar patterns. Consequently, we define our spatiotemporal vector representation $V$ as

\begin{equation}\label{eq:query}
    V \left( \hat{u}, \,\hat{d},\, \hat{t} \right) \, = \, \left(lat^{\hat{u}}_{\hat{d},\,\hat{t}},\, lon^{\hat{u}}_{\hat{d},\,\hat{t}},\, \alpha \cdot cos \frac{2\pi \hat{t}}{T},\, \alpha \cdot cos \frac{2\pi \hat{t}}{T} \right)
\end{equation}

where $\alpha$ is the coefficient controlling the relative weights between continuity and periodicity. A $k$-d tree is built in the offline phase to prepare for the online searching of the $N$-nearest neighbors as candidates.

\subsection{Online Prediction}
The online prediction problem can be formulated as:
\begingroup
\small
\begin{multline}
    p\left(Trn^u[t:\,t + \Delta T] \, \bigg|\, TR[t - \Delta T:\,t] \right) =\\
    \sum_{C} p\left( Trn^u[t:\,t + \Delta T] \, \bigg|\, C,\, Tr^u[t]\,\right) p\left(\, C \, \bigg|\, TRC[t - \Delta T:\,t] \right)
    \label{eq:online_prediction}
\end{multline}
\endgroup

where the raw trajectories set from all user $TR[t - \Delta T:\, t] = \left\lbrace Tr^{u}[t - \Delta T\,:\, t] \,\bigg|\, u \in U\right\rbrace$, and $TR$ can be transformed to $TRC$ in a deterministic manner. $C$ denotes the $\Delta T$-ahead destination cluster $Trc^u_{t + \Delta T}$. Note that the two terms represent our two-stage prediction, respectively.

1) $p\left(\, C \, \bigg|\, TRC[t - \Delta T:\,t] \right)$ represents the probability distribution predicted by the cluster-level predictor. This stage is simple, and we have introduced the pre-computing strategy in Section \ref{sec:pretrain}.

2) Because the intrinsic structure of $Trn^u[t:\,t + \Delta T]$ is very hard to model, we approximate the distribution $p\left( Trn^u[t:\,t + \Delta T] \, \bigg|\, C,\, Tr^u[t]\,\right)$ via a prediction-by-retrieval approach. This can be done using a four-step process in a Monte Carlo Markov chain.
\begin{itemize}
\item Determine the shard of historical trajectory database $DB[C]$ by sampling destination cluster prediction $C$.
\item Calculate the query vector $V$ similar to Equation \ref{eq:query} from $Tr^u[t]$, and search the $K$ nearest neighbors $\left\lbrace (Trn_k, \,\epsilon_k) \,\bigg|\, Trn_k \in DB[C] \right\rbrace$ where $k = 1,\dots,K$ with distance $\epsilon_k$.
\item Weighting the probability of candidates $\left\lbrace Trn_k \right\rbrace$ by their distances: $p\left(Trn_k\right) = \frac{exp(-\beta \epsilon_k)}{\sum_j exp(-\beta \epsilon_j)} $, where $\beta$ is the parameter controlling the sampling temperature.
\item Candidate is drawn from $p\left(Trn_k\right)$. Minor changes such as changing the user ID to $u$ and updating the time to the current time are applied before exporting our fine-grained prediction results.
\end{itemize}

Note that, because all the trajectories in the historical database are fine-grained trajectories, our predicted trajectories, which are retrieved from the database as the prototypes, are also fine-grained trajectories. In other words, we avoid the heavy computation and complex modeling asscociated with generating a fine-grained trajectory by retrieving fine-grained trajectories in the historical database, where the sampling weight is determined by the cluster-level prediction.
\begin{figure*}[!h]
\centering
\includegraphics[width=0.82\textwidth]{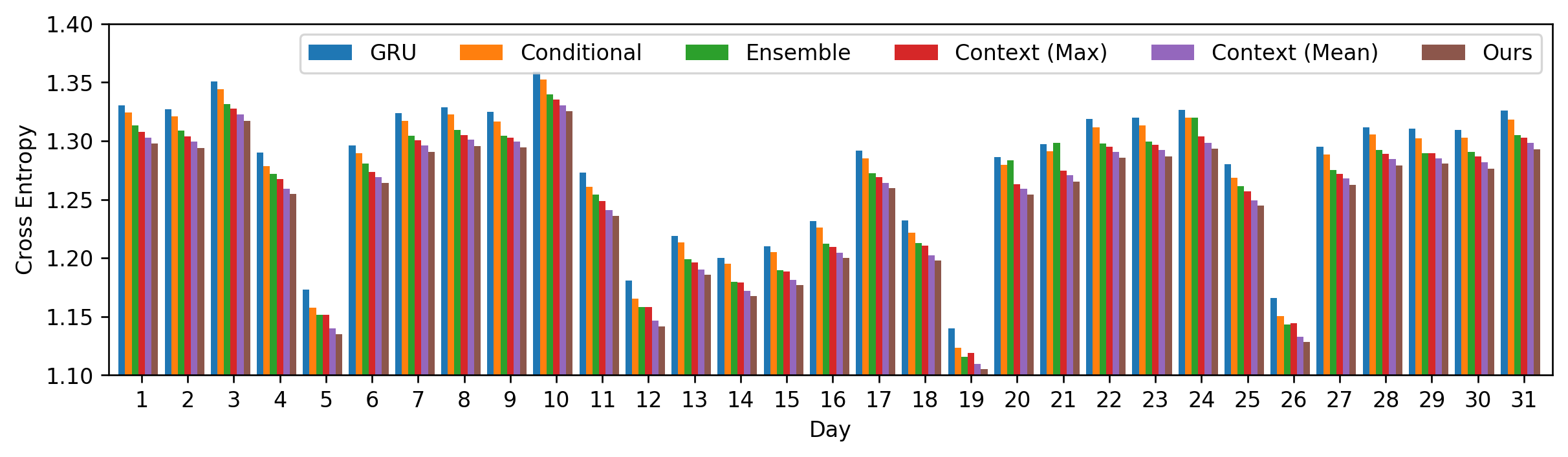}
\vspace{-2mm}
\caption{Evaluation on cluster-level prediction on different days}
\label{fig:cross_entropy_evaluation}
\end{figure*}
\section{Experimental results}
\subsection{Data}
In this paper, we use a dataset "Konzatsu-Tokei (R)" data. “Konzatsu-Tokei (R)" Data refers to people flows data collected by individual location data sent from mobile phone under users' consent, through applications provided by NTT DOCOMO, INC. Those data are processed collectively and statistically in order to conceal the private information. Original location data is GPS data (latitude, longitude) sent in about every a minimum period of 5 minutes and does not include the information to specify individual. Some applications such as “docomo map navi" service (map navi and local guide).

We use two months of data (from 2012.6.1 to 2012.07.31) for training/validating the cluster-level predictor and building the fine-grained historical trajectory database, and one month data (from 2012.08.01 to 2012.08.31) for evaluation. We cropped the data in Kanto area (covering Tokyo metropolitan area), with an average number of about 220K user IDs. For the sake of protecting user's privacy as well as efficiency, we do not store users' ID for long-observation and thus we are more focused on the movement distribution at the large scale while a long dependency of the user's mobility \cite{Feng:2018:DPH:3178876.3186058} is not taken into consideration.

\subsection{Evaluation on Cluster-level Prediction}
We choose five baseline methods to evaluate our cluster-level prediction.
\begin{itemize}
    \item \textbf{GRU} is the predictor that is defined in Equation \ref{eq:base_prediction};
    \item \textbf{Conditional} is an extension to GRU, which uses day-of-week as auxiliary data to distinguish between weekdays and weekends;
    \item \textbf{Ensemble} is based on the method in \cite{fan2018online}, which trains an independent predictor for each single day in the training dataset, and use an gating function to fuse these predictors in an adaptive way. In this experiment, we pre-train 14 predictors from Jun 1 to Jun 14;
    \item \textbf{Context (Mean)} excludes the top GRU layer for crowd context. Thus, the ability of modeling sequential pattern of the crowd context is weakened.
    \item \textbf{Context (Max)} shares the same network structure with ``Context (Mean)", but uses $Max$ as a pooling function.
\end{itemize}

We use cross entropy (a lower value that indicates a better performance) for evaluation. As is shown in Fig. \ref{fig:cross_entropy_evaluation}, our cluster-level predictor achieves the best performance in our test data. The non-adaptive predictor GRU performs worst because it completely ignores the different mobility patterns of users on different days. ``Conditional" is capable of distinguishing between weekdays and holidays, but fails to capture the subtle difference between different ``weekdays" or ``holidays" from crowd context. ``Mean" pooling functions better than ``Max" pooling, because the average operation is better at preserving the crowd proportion information than maxing out. ``Ensemble" is limited by the number of pre-trained predictors. We observed a decrease of cross entropy if we add more predictors; however the speed drops significantly for several predictors (about 5 times slower than our proposed method if we use 14 predictors), making it less practical for an online prediction system.

\subsection{Evaluation on Fine-grained Prediction}
\begin{figure}[h!]
\centering
\includegraphics[width=\linewidth]{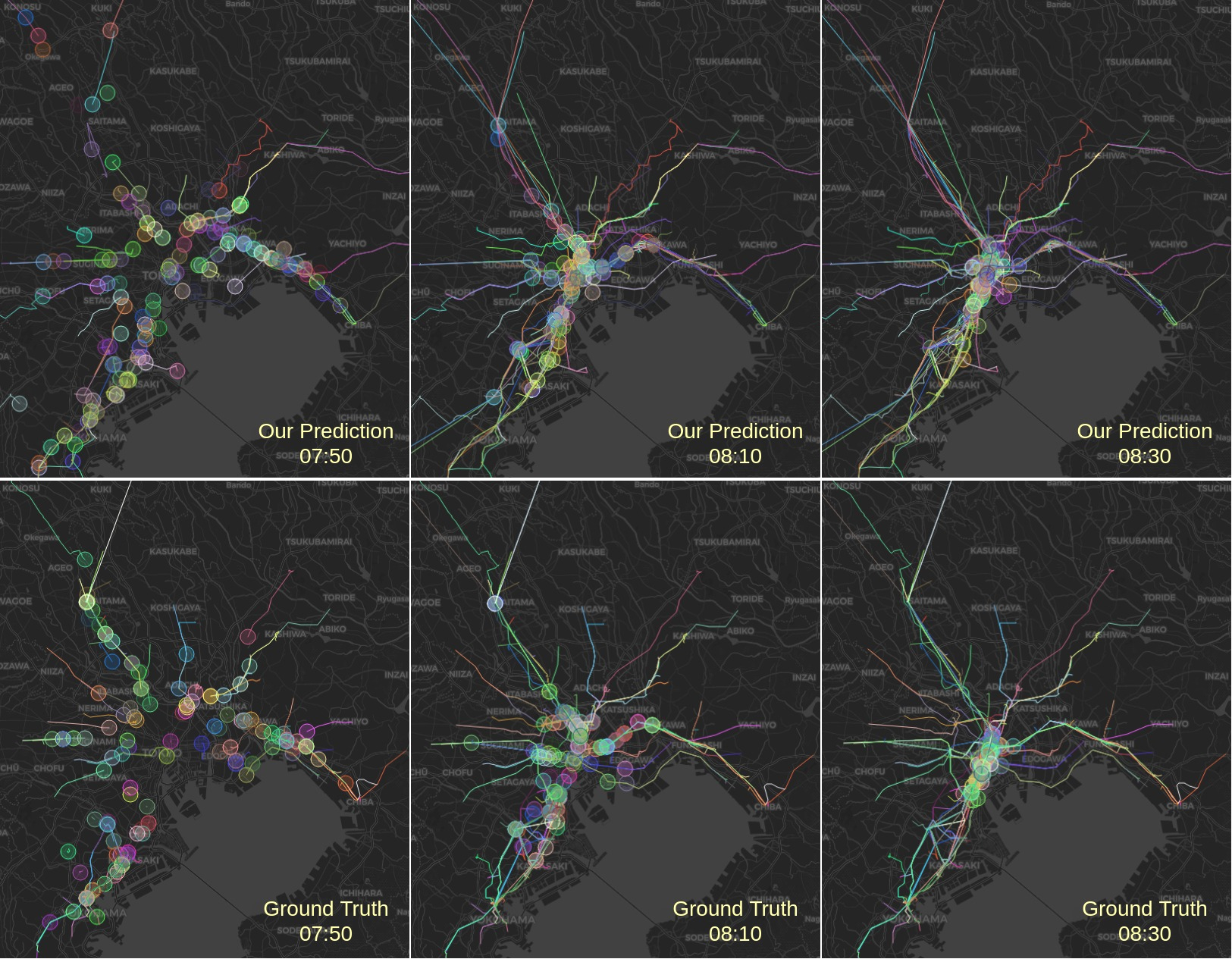}
\caption{Gathering pattern of Tokyo station at morning rush hour}
\label{fig:gathering_pattern}
\end{figure}
\begin{figure*}[h!]
\centering
\includegraphics[width=0.93\textwidth]{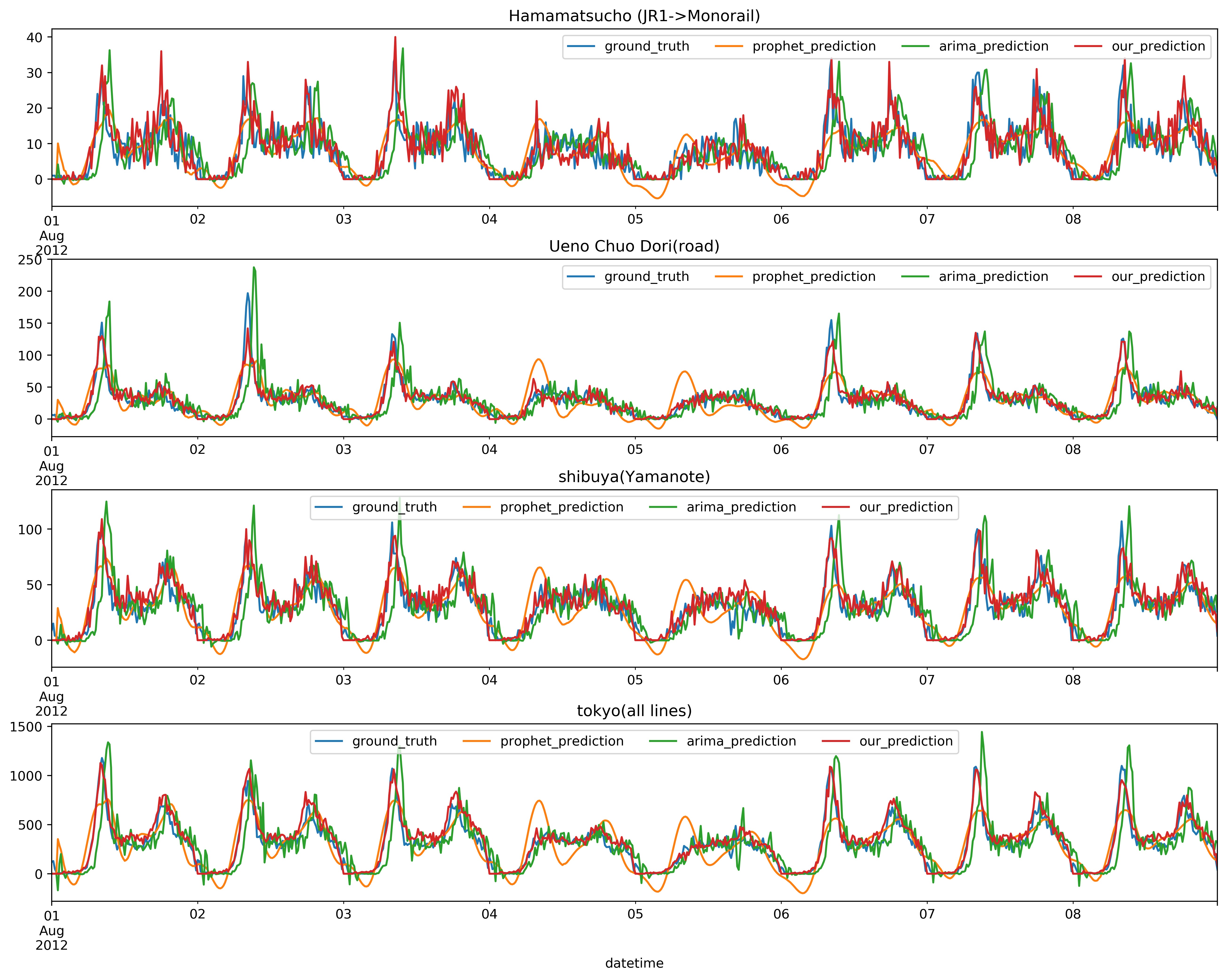}
\vspace{-6mm}
\caption{Transportation node-level aggregation evaluation of our prediction results.}
\label{fig:aggregated_prediction_results}
\end{figure*}
As shown in Fig. \ref{fig:gathering_pattern}, we visualize our prediction results at time 07:30 on a weekday morning (2012.08.01 Wed) to qualitatively evaluate how well the gathering pattern in the morning rush hour for Tokyo station is predicted. 1h-ahead fine-grained trajectories are predicted and compared with ground truth trajectories for those trajectories passing Tokyo station during 08:15-08:30. We take a snapshot of the trajectories every 20 minutes. Consequently, it is observed that our proposed method can predict how people will gather at Tokyo station (e.g., the source distribution of the gathering people and which route they will take) well.
In Fig. \ref{fig:aggregated_prediction_results}, we selected four types of aggregated analysis: 1) number of transfer, 2) traffic volume on a particular road segment, 3) number of passengers using the station on a particular line, and 4) the total number of passengers for all lines in an interchange. We compared our 1h-ahead prediction with the most widely used time-series prediction methods, namely ARIMA\cite{10.5555/574978} and prophet\cite{DBLP:journals/peerjpre/TaylorL17} from 2012.08.01 to 2012.08.08. We can see from Fig. \ref{fig:aggregated_prediction_results} that our fine-grained prediction can predict the sharp peaks of rush hours accurately and switch between the weekday and weekend state via crowd context automatically, which are difficult for traditional time series methods. A more comprehensive quantitative evaluation is given in Table \ref{tab:quantitative}. In general, our proposed method achieves a higher accuracy, especially for "all lines" which is less affected by random noise caused by few observations. Note that our predictor can predict more accurately, and is more informative fo the fine-grained trajectories compared with baseline models.

\begin{table*}[h!]
\centering
\caption{Evaluation on Fine-grained Prediction}
\label{tab:quantitative}
\begin{tabular}{|l|c|c|c|c|}
\hline & \multicolumn{4}{c|}{RMSE / MAE / MAPE} \\ \hline
& Ours          & GRU   & ARIMA & Prophet\\ \hline
Nihonbashi                & \textbf{9.8 / 6.8 / 0.2379}  & 10.0 / 7.1 / 0.2458  & 17.7 / 11.9 / 0.4153  & 12.3 / 9.3 / 0.3233\\ \hline
Nishi Ikebukuro           & 11.9 / \textbf{8.8} / \textbf{0.2563}  & 13.6 / 10.1 / 0.2955   & 16.6 / 12.0 / 0.3500 & \textbf{11.5} / 8.9 / 0.2594 \\ \hline
14 national road          & \textbf{9.0 / 6.6 / 0.2517}  & 11.5 / 8.2 / 0.3141  & 18.7 / 11.9 / 0.4549 & 12.7 / 9.4 / 0.3621\\ \hline
Ueno Chuo Dori            & \textbf{8.7 / 5.9 / 0.2211}  & 11.7 / 7.7 / 0.2861 & 21.3 / 12.4 / 0.4649 & 14.9 / 10.2 / 0.3814  \\ \hline
20 national road          & \textbf{10.2 / 7.4 / 0.2871} & 11.8 / 8.6 / 0.3340 & 15.1 / 10.3 / 0.4017 & 10.6 / 7.9 / 0.3070\\ \hline
Shinjuku(Yamanote)  & \textbf{17.2 / 12.3 / 0.2159} & 20.6 / 14.9 / 0.2616  & 32.7 / 22.0 / 0.3856  & 22.1 / 16.8 / 0.2948  \\ \hline
Tokyo(Yamanote)     & \textbf{12.8 / 8.8 / 0.2071} & 14.6 / 10.2 / 0.2407 & 27.6 / 17.7 / 0.4170  & 18.8 / 14.1 / 0.3316  \\ \hline
Shibuya(Yamanote)   & \textbf{9.8 / 6.9 / 0.2554}  & 10.9 / 7.7 / 0.2830 & 17.4 / 11.7 / 0.4301  & 11.7 / 8.7 / 0.3214  \\ \hline
Ueno(Yamanote)      & \textbf{12.6 / 8.9 / 0.2038} & 15.1 / 10.9 / 0.2512 & 26.5 / 17.7 / 0.4072 & 17.6 / 13.4 / 0.3078  \\ \hline
Akihabara(Yamanote) & \textbf{9.3 / 6.5 / 0.2876}  & 10.7 / 7.7 / 0.3391 & 14.6 / 10.0 / 0.4398  & 10.1 / 7.7 / 0.3396  \\ \hline
Shinagawa(Yamanote) & \textbf{14.7 / 9.6 / 0.1986} & 16.9 / 11.2 / 0.2321 & 34.5 / 21.9 / 0.4514  & 23.6 / 17.2 / 0.3552  \\ \hline
Takadanobaba(Seibu) & \textbf{7.1 / 5.1 / 0.3135} & 8.0 / 5.7 / 0.3518  & 11.2 / 7.8 / 0.4816  & 7.6 / 5.9 / 0.3632  \\ \hline
Ueno(Ginza) & \textbf{7.4 / 5.2 / 0.2840} & 8.6 / 6.1 / 0.3333 & 12.9 / 8.5 / 0.4657  & 8.8 / 6.6 / 0.3608  \\ \hline
Ueno(Joban) & \textbf{4.2 / 2.7 / 0.3923} & 4.8 / 3.1 / 0.4391  & 8.2 / 4.6 / 0.6594  & 5.4 / 3.7 / 0.5300  \\ \hline
Shinagawa(Shinkansen) & \textbf{17.7 / 12.6 / 0.2993} & 20.0 / 14.5 / 0.3448 & 29.4 / 18.7 / 0.4453  & 20.1 / 14.8 / 0.3511  \\ \hline
Shinjuku(all lines)      & \textbf{87.1 / 64.9 / 0.1950} & 111.3 / 82.8 / 0.2487 & 173.5 / 116.0 / 0.3483 & 122.9 / 94.0 / 0.2822  \\ \hline
Tokyo(all lines)         & \textbf{72.9 / 51.1 / 0.1794} & 92.2 / 66.7 / 0.2342  & 172.2 / 108.4 / 0.3806 & 121.9 / 90.9 / 0.3191  \\ \hline
Shibuya(all lines)       & \textbf{50.1 / 37.0 / 0.2245} & 59.9 / 44.3 / 0.2693  & 86.0 / 58.7 / 0.3567  & 59.8 / 45.4 / 0.2759  \\ \hline
Ueno(all lines)          & \textbf{56.8 / 40.7 / 0.1819} & 75.2 / 54.2 / 0.2424  & 138.0 / 87.2 / 0.3896 & 96.6 / 72.1 / 0.3222  \\ \hline
Akihabara(all lines)     & \textbf{30.2 / 21.6 / 0.2435} & 39.0 / 28.3 / 0.3188  & 61.1 / 38.5 / 0.4333  & 43.2 / 31.6 / 0.3551  \\ \hline
Ikebukuro(all lines)     & \textbf{65.7 / 49.2 / 0.2158} & 84.8 / 61.5 / 0.3533  & 118.2 / 80.5 / 0.3533 & 82.6 / 63.4 / 0.2782  \\ \hline
Shinagawa(all lines)     & \textbf{51.9 / 36.5 / 0.1975} & 68.2 / 48.7 / 0.2633  & 128.8 / 80.1 / 0.4333 & 91.4 / 66.9 / 0.3619  \\ \hline
Nishinippori(Metro->JR)     & \textbf{4.2 / 2.8 / 0.4124} & 4.3 / 2.8 / 0.4146  & 6.4 / 4.1 / 0.5990 & 4.3 / 3.1 / 0.4561  \\ \hline
Shinyokohama(JR->SKS)     & 3.6 / 2.5 / 0.4398 & 3.6 / 2.5 / 0.4470  & 4.5 / 3.2 / 0.5650 & \textbf{3.2} / \textbf{2.4} / \textbf{0.4237}  \\ \hline
Hamamatsucho(JR->MR)     & 4.4 / \textbf{3.1} / \textbf{0.3972} & 4.6 / 3.2 / 0.4151  & 6.2 / 4.1 / 0.5357 & \textbf{4.3} / 3.2 / 0.4178  \\ \hline
\end{tabular}
\end{table*}
\begin{figure}[h!]
\centering
\includegraphics[width=\linewidth]{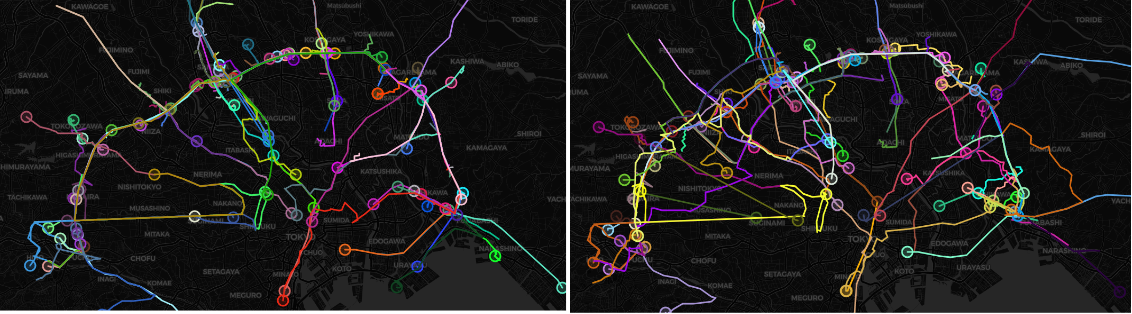}
\caption{Prediction of the users that are affected by the stop of the Musashino line (left) and simulation of the alternative routes without using Musashino line (right).}
\label{fig:simulation}
\end{figure}
\subsection{Simulation}
Note that our proposed prediction system for mobility digital twins can not only predict the future movements of the users based on the current urban status, but also predicting the trajectories responding to different conditions by filtering or augmenting the historical database with respect to specific simulation tasks. For example, we can easily identify those users that are potentially affected by the suspension of the Musashino line (the Tokyo unclosed outer ring line) from our fine-grained trajectory prediction results, as shown on the left of Fig. \ref{fig:simulation}. Assuming the users will not change their moving destinations, we can easily generate the simulated future movements of those affected users if the Musashino line is out-of-service by filtering out all those candidates trajectories in the historical trajectory database that travels with Musashino line. As shown on the right of Figure \ref{fig:simulation}, alternative routes are found from the historical trajectory database with the consideration of the frequencies, the current location, and the destination region. Such analysis and simulation are not well-supported by most of existing aggregated level or destination-only coarse mobility prediction methods.

\subsection{System Implementation and Time Efficiency}
We deployed our algorithm on a deep learning workstation with Intel Xeon E5-2690v4, 2 x TitanX Pascal 12GB GDDR5X, 128GB and 1.2TB Intel® NVMe SSD DC P3600 Series. The algorithm was implemented in Python, except the trajectory interpolation and map-matching part were implemented in Java. We utilized the deep learning framework PyTorch 1.3.1 \footnote{\url{https://pytorch.org/}} to construct the cluster-level predictor, and key-value storage library levelDB 1.20 \footnote{\url{https://github.com/google/leveldb}} for building and online querying the fine-grained history trajectory database.

We measure the time efficiency of the three steps in our online prediction phase. From Table \ref{tab:efficiency}, we can see that fine-grained 1h-ahead trajectory prediction for about 220K users at a metropolitan scale takes about 2 min, which is a reasonable time latency for a practical system. Note that, except computing the crowd context, cluster-level and node-level prediction are conducted independently on each user. Thus a parallel acceleration strategy can be expected to achieve a desirable acceleration in our future work.
\begin{table}[]
\centering
\caption{Time efficiency}
\label{tab:efficiency}
\begin{tabular}{|l|l|l|}
\hline
Context (s)   & Cluster-level (s) & Node-level (s) \\ \hline
0.3726±0.0119 & 21.873±1.216     & 112.35±19.04          \\ \hline
\end{tabular}
\end{table}
\section{Related work}

Digital twin \cite{tao2018digital}, which makes a replication of the physical world in the digital world, has drawn increasing attention in recent years, especially in the research of smart city \cite{deng2021systematic, wang2022mobility, clemen2021multi, seto2020constructing}. However, most studies on digital twin for smart city focus more on the real-time sensing and interactive visualization, while the real-time trajectory prediction algorithm that is suitable for the mobility replica in the digital twin is relatively under-explored.

Most existing studies predict human mobility at a small scale \cite{song2010online, alahi2016social, zhao2018prediction}, aggregated level \cite{lv2014traffic, RNN, CHEN201859, yao2018deep, Konishi:2016:CCI:2971648.2971718, zhang2017deep, s18103431, lin2019deepstn+} or coarse-level \cite{fan2018online, song2016deeptransport, Feng:2018:DPH:3178876.3186058, zhao2018prediction}. Compared with our fine-grained predictor, such coarse predictions are insufficient to support higher-level transportation predictive analysis, such as transportation transfer prediction (e.g., transfer between railway lines or to other transportation modes).

To predict the fine-grained level human mobility, we need to generate a realistic trajectory, which is a long timestamped sequence, on the transportation network. \cite{jiang2018deepurbanmomentum} applied the seq2seq model to human trajectories and predicted future movements at the coarse level (only a few steps with continuous location representation), without considering the transportation network. \cite{10.1145/3287064} predicted the rest-of-the-day trajectory of the user in a predicting-by-retrieving paradigm, which is similar to the node-level prediction phase in our work. However, only the user's historical trajectories are considered, and only the dynamic time warping distances are calculated for measuring the similarity between the most recent trajectory and historical trajectories. Thus, our work is more suitable for predicting human mobility at a metropolitan scale.

Moreover, the various nodes/links in the network makes the SoftMax layer very difficult to train. \cite{DBLP:conf/icml/GraveJCGJ17} uses clustering on the classes based on the frequency, and can accept a much larger number of classes. However, a transportation network within a metropolitan area has a much larger number of unique nodes/links (e.g., in our application we have $10^6$ unique nodes/links in the Greater Tokyo area), making it hard to model. An alternative that circumvents the above problem is to use ranking models \cite{Kenter:2017:NNI:3077136.3082062}, which well preserves the structure information. If our expected output text is most probably included, or can be approximated from the database, ranking models will significantly outperform generative models in terms of the quality of generated sequences. In this study, we utilize the second method, but approximate distribution over historical trajectories in a feasible manner to avoid bias in the aggregated traffic flow.

Many existing predictors for individual users emphasize predicting the regularities of human mobility \cite{gonzalez2008understanding, Feng:2018:DPH:3178876.3186058, zhao2018prediction, rossi2019modelling}. This may lead to a drift when the crowd trend changes. Some studies have explored methods considering crowd conformity: \cite{10.1145/2632048.2632063} predicted the user attendance of an event by leveraging both local and global historical data from all users; \cite{yang2020location} proposed a flashback on hidden states of recurrent neural network to model the periodicity of spatiotemporal contexts of location-based service user's sparse trajectory. \cite{10.1145/2783258.2783350} modeled the regularity and conformity in human mobility, where conformity is the pattern in which some users follow others. In this study, conformity is modeled via crowd context implicitly. \cite{fan2018online} trained an independent component predictor on each day in the training set, and performed online adaptive learning to leverage the crowd trend information. This method can make a good prediction for both regular and irregular mobility; however, numerous component predictors make the the prediction very inefficient (time and memory). Social LSTM \cite{alahi2016social} designs a social pooling that is capable of integrating the mobility of a crowd. This idea is the most relevant to our cluster-level predictor with crowd context; however, this study aims to solve the human mobility prediction for a surveillance scene. Inspired by Social LSTM and some following studies \cite{vemula2018social}, we design a batchwise mean pooling method to capture the crowd context at each time step, and a context GRU to model the sequential pattern in the crowd context, which is more suitable for large-scale mobility prediction.

\section{Conclusion}
This study proposes a novel two-stage fine-grained human mobility prediction method for predicting the mobility replica for the mobilty digital twin. Crowd context, that describes the current state of the mobility replica, is considered to make the predictor more flexible to perceive the current crowd trend information, and a predict-by-retrieval method is proposed to predict fine-grained trajectories in a practical time.

We also note some limitations of the current work. 1) Our crowd context is described in a global scale, while local irregularity can be ignored by our global crowd contexts if it is less significant to be aware at the metropolitan scale. Thus, a fusion of global and local crowd context is considered as a promising future direction. 2) We currently implement our online prediction system on a single machine in which little optimization of time efficiency is considered. As we can see from Table \ref{tab:efficiency}, the time efficiency bottleneck of current system is within node-level prediction, which takes up about 84\% of the total elapsed time. The interdependence of different users are mainly modeled in the crowd context computing phase, which takes only about 0.3\% of the total elapsed time, while cluster-level and node-level prediction is conducted on each user independently. Thus, parallel computing is very promising in accelerating our prediction significantly.

\bibliographystyle{ACM-Reference-Format}
\bibliography{sample-base}


\begin{thebibliography}{32}


\ifx \showCODEN    \undefined \def \showCODEN     #1{\unskip}     \fi
\ifx \showDOI      \undefined \def \showDOI       #1{#1}\fi
\ifx \showISBNx    \undefined \def \showISBNx     #1{\unskip}     \fi
\ifx \showISBNxiii \undefined \def \showISBNxiii  #1{\unskip}     \fi
\ifx \showISSN     \undefined \def \showISSN      #1{\unskip}     \fi
\ifx \showLCCN     \undefined \def \showLCCN      #1{\unskip}     \fi
\ifx \shownote     \undefined \def \shownote      #1{#1}          \fi
\ifx \showarticletitle \undefined \def \showarticletitle #1{#1}   \fi
\ifx \showURL      \undefined \def \showURL       {\relax}        \fi
\providecommand\bibfield[2]{#2}
\providecommand\bibinfo[2]{#2}
\providecommand\natexlab[1]{#1}
\providecommand\showeprint[2][]{arXiv:#2}

\bibitem[\protect\citeauthoryear{Alahi, Goel, Ramanathan, Robicquet, Fei-Fei,
  and Savarese}{Alahi et~al\mbox{.}}{2016}]%
        {alahi2016social}
\bibfield{author}{\bibinfo{person}{Alexandre Alahi}, \bibinfo{person}{Kratarth
  Goel}, \bibinfo{person}{Vignesh Ramanathan}, \bibinfo{person}{Alexandre
  Robicquet}, \bibinfo{person}{Li Fei-Fei}, {and} \bibinfo{person}{Silvio
  Savarese}.} \bibinfo{year}{2016}\natexlab{}.
\newblock \showarticletitle{Social lstm: Human trajectory prediction in crowded
  spaces}. In \bibinfo{booktitle}{\emph{Proceedings of the IEEE Conference on
  Computer Vision and Pattern Recognition}}. \bibinfo{pages}{961--971}.
\newblock


\bibitem[\protect\citeauthoryear{Box and Jenkins}{Box and Jenkins}{1990}]%
        {10.5555/574978}
\bibfield{author}{\bibinfo{person}{George Edward~Pelham Box} {and}
  \bibinfo{person}{Gwilym Jenkins}.} \bibinfo{year}{1990}\natexlab{}.
\newblock \bibinfo{booktitle}{\emph{Time Series Analysis, Forecasting and
  Control}}.
\newblock \bibinfo{publisher}{Holden-Day, Inc.}, \bibinfo{address}{USA}.
\newblock
\showISBNx{0816211043}


\bibitem[\protect\citeauthoryear{Chen, Yang, Zhang, Wang, Li, and Nguyen}{Chen
  et~al\mbox{.}}{2018}]%
        {CHEN201859}
\bibfield{author}{\bibinfo{person}{Longbiao Chen}, \bibinfo{person}{Dingqi
  Yang}, \bibinfo{person}{Daqing Zhang}, \bibinfo{person}{Cheng Wang},
  \bibinfo{person}{Jonathan Li}, {and} \bibinfo{person}{Thi-Mai-Trang Nguyen}.}
  \bibinfo{year}{2018}\natexlab{}.
\newblock \showarticletitle{Deep mobile traffic forecast and complementary base
  station clustering for C-RAN optimization}.
\newblock \bibinfo{journal}{\emph{Journal of Network and Computer
  Applications}}  \bibinfo{volume}{121} (\bibinfo{year}{2018}),
  \bibinfo{pages}{59 -- 69}.
\newblock
\showISSN{1084-8045}
\urldef\tempurl%
\url{https://doi.org/10.1016/j.jnca.2018.07.015}
\showDOI{\tempurl}


\bibitem[\protect\citeauthoryear{Clemen, Ahmady-Moghaddam, Lenfers, Ocker,
  Osterholz, Str{\"o}bele, and Glake}{Clemen et~al\mbox{.}}{2021}]%
        {clemen2021multi}
\bibfield{author}{\bibinfo{person}{Thomas Clemen}, \bibinfo{person}{Nima
  Ahmady-Moghaddam}, \bibinfo{person}{Ulfia~A Lenfers},
  \bibinfo{person}{Florian Ocker}, \bibinfo{person}{Daniel Osterholz},
  \bibinfo{person}{Jonathan Str{\"o}bele}, {and} \bibinfo{person}{Daniel
  Glake}.} \bibinfo{year}{2021}\natexlab{}.
\newblock \showarticletitle{Multi-agent systems and digital twins for smarter
  cities}. In \bibinfo{booktitle}{\emph{Proceedings of the 2021 ACM SIGSIM
  Conference on Principles of Advanced Discrete Simulation}}.
  \bibinfo{pages}{45--55}.
\newblock


\bibitem[\protect\citeauthoryear{Deng, Zhang, and Shen}{Deng
  et~al\mbox{.}}{2021}]%
        {deng2021systematic}
\bibfield{author}{\bibinfo{person}{Tianhu Deng}, \bibinfo{person}{Keren Zhang},
  {and} \bibinfo{person}{Zuo-Jun~Max Shen}.} \bibinfo{year}{2021}\natexlab{}.
\newblock \showarticletitle{A systematic review of a digital twin city: A new
  pattern of urban governance toward smart cities}.
\newblock \bibinfo{journal}{\emph{Journal of Management Science and
  Engineering}} \bibinfo{volume}{6}, \bibinfo{number}{2}
  (\bibinfo{year}{2021}), \bibinfo{pages}{125--134}.
\newblock


\bibitem[\protect\citeauthoryear{Du, Yu, Mei, Wang, Wang, and Guo}{Du
  et~al\mbox{.}}{2014}]%
        {10.1145/2632048.2632063}
\bibfield{author}{\bibinfo{person}{Rong Du}, \bibinfo{person}{Zhiwen Yu},
  \bibinfo{person}{Tao Mei}, \bibinfo{person}{Zhitao Wang},
  \bibinfo{person}{Zhu Wang}, {and} \bibinfo{person}{Bin Guo}.}
  \bibinfo{year}{2014}\natexlab{}.
\newblock \showarticletitle{Predicting Activity Attendance in Event-Based
  Social Networks: Content, Context and Social Influence}. In
  \bibinfo{booktitle}{\emph{Proceedings of the 2014 ACM International Joint
  Conference on Pervasive and Ubiquitous Computing}} (Seattle, Washington)
  \emph{(\bibinfo{series}{UbiComp ’14})}. \bibinfo{publisher}{Association for
  Computing Machinery}, \bibinfo{address}{New York, NY, USA},
  \bibinfo{pages}{425–434}.
\newblock
\showISBNx{9781450329682}


\bibitem[\protect\citeauthoryear{Fan, Song, Xia, Jiang, Shibasaki, and
  Sakuramachi}{Fan et~al\mbox{.}}{2018}]%
        {fan2018online}
\bibfield{author}{\bibinfo{person}{Zipei Fan}, \bibinfo{person}{Xuan Song},
  \bibinfo{person}{Tianqi Xia}, \bibinfo{person}{Renhe Jiang},
  \bibinfo{person}{Ryosuke Shibasaki}, {and} \bibinfo{person}{Ritsu
  Sakuramachi}.} \bibinfo{year}{2018}\natexlab{}.
\newblock \showarticletitle{Online deep ensemble learning for predicting
  citywide human mobility}.
\newblock \bibinfo{journal}{\emph{Proceedings of the ACM on Interactive,
  Mobile, Wearable and Ubiquitous Technologies}} \bibinfo{volume}{2},
  \bibinfo{number}{3} (\bibinfo{year}{2018}), \bibinfo{pages}{1--21}.
\newblock


\bibitem[\protect\citeauthoryear{Feng, Li, Xu, and Jin}{Feng
  et~al\mbox{.}}{2018a}]%
        {s18103431}
\bibfield{author}{\bibinfo{person}{Jie Feng}, \bibinfo{person}{Yong Li},
  \bibinfo{person}{Fengli Xu}, {and} \bibinfo{person}{Depeng Jin}.}
  \bibinfo{year}{2018}\natexlab{a}.
\newblock \showarticletitle{A Bimodal Model to Estimate Dynamic Metropolitan
  Population by Mobile Phone Data}.
\newblock \bibinfo{journal}{\emph{Sensors}} \bibinfo{volume}{18},
  \bibinfo{number}{10} (\bibinfo{year}{2018}).
\newblock
\showISSN{1424-8220}


\bibitem[\protect\citeauthoryear{Feng, Li, Zhang, Sun, Meng, Guo, and Jin}{Feng
  et~al\mbox{.}}{2018b}]%
        {Feng:2018:DPH:3178876.3186058}
\bibfield{author}{\bibinfo{person}{Jie Feng}, \bibinfo{person}{Yong Li},
  \bibinfo{person}{Chao Zhang}, \bibinfo{person}{Funing Sun},
  \bibinfo{person}{Fanchao Meng}, \bibinfo{person}{Ang Guo}, {and}
  \bibinfo{person}{Depeng Jin}.} \bibinfo{year}{2018}\natexlab{b}.
\newblock \showarticletitle{DeepMove: Predicting Human Mobility with
  Attentional Recurrent Networks}. In \bibinfo{booktitle}{\emph{Proceedings of
  the 2018 World Wide Web Conference}} (Lyon, France)
  \emph{(\bibinfo{series}{WWW '18})}. \bibinfo{pages}{1459--1468}.
\newblock
\showISBNx{978-1-4503-5639-8}


\bibitem[\protect\citeauthoryear{Gonzalez, Hidalgo, and Barabasi}{Gonzalez
  et~al\mbox{.}}{2008}]%
        {gonzalez2008understanding}
\bibfield{author}{\bibinfo{person}{Marta~C Gonzalez}, \bibinfo{person}{Cesar~A
  Hidalgo}, {and} \bibinfo{person}{Albert-Laszlo Barabasi}.}
  \bibinfo{year}{2008}\natexlab{}.
\newblock \showarticletitle{Understanding individual human mobility patterns}.
\newblock \bibinfo{journal}{\emph{nature}} \bibinfo{volume}{453},
  \bibinfo{number}{7196} (\bibinfo{year}{2008}), \bibinfo{pages}{779--782}.
\newblock


\bibitem[\protect\citeauthoryear{Grave, Joulin, Ciss{\'{e}}, Grangier, and
  J{\'{e}}gou}{Grave et~al\mbox{.}}{2017}]%
        {DBLP:conf/icml/GraveJCGJ17}
\bibfield{author}{\bibinfo{person}{Edouard Grave}, \bibinfo{person}{Armand
  Joulin}, \bibinfo{person}{Moustapha Ciss{\'{e}}}, \bibinfo{person}{David
  Grangier}, {and} \bibinfo{person}{Herv{\'{e}} J{\'{e}}gou}.}
  \bibinfo{year}{2017}\natexlab{}.
\newblock \showarticletitle{Efficient softmax approximation for GPUs}. In
  \bibinfo{booktitle}{\emph{Proceedings of the 34th International Conference on
  Machine Learning, {ICML} 2017, Sydney, NSW, Australia, 6-11 August 2017}}.
  \bibinfo{pages}{1302--1310}.
\newblock
\urldef\tempurl%
\url{http://proceedings.mlr.press/v70/grave17a.html}
\showURL{%
\tempurl}


\bibitem[\protect\citeauthoryear{Jiang, Song, Fan, Xia, Chen, Miyazawa, and
  Shibasaki}{Jiang et~al\mbox{.}}{2018}]%
        {jiang2018deepurbanmomentum}
\bibfield{author}{\bibinfo{person}{Renhe Jiang}, \bibinfo{person}{Xuan Song},
  \bibinfo{person}{Zipei Fan}, \bibinfo{person}{Tianqi Xia},
  \bibinfo{person}{Quanjun Chen}, \bibinfo{person}{Satoshi Miyazawa}, {and}
  \bibinfo{person}{Ryosuke Shibasaki}.} \bibinfo{year}{2018}\natexlab{}.
\newblock \showarticletitle{DeepUrbanMomentum: An Online Deep-Learning System
  for Short-Term Urban Mobility Prediction.}. In
  \bibinfo{booktitle}{\emph{AAAI}}.
\newblock


\bibitem[\protect\citeauthoryear{Kenter, Borisov, Van~Gysel, Dehghani,
  de~Rijke, and Mitra}{Kenter et~al\mbox{.}}{2017}]%
        {Kenter:2017:NNI:3077136.3082062}
\bibfield{author}{\bibinfo{person}{Tom Kenter}, \bibinfo{person}{Alexey
  Borisov}, \bibinfo{person}{Christophe Van~Gysel}, \bibinfo{person}{Mostafa
  Dehghani}, \bibinfo{person}{Maarten de Rijke}, {and} \bibinfo{person}{Bhaskar
  Mitra}.} \bibinfo{year}{2017}\natexlab{}.
\newblock \showarticletitle{Neural Networks for Information Retrieval}. In
  \bibinfo{booktitle}{\emph{Proceedings of the 40th International ACM SIGIR
  Conference on Research and Development in Information Retrieval}} (Shinjuku,
  Tokyo, Japan) \emph{(\bibinfo{series}{SIGIR '17})}. \bibinfo{publisher}{ACM},
  \bibinfo{address}{New York, NY, USA}, \bibinfo{pages}{1403--1406}.
\newblock
\showISBNx{978-1-4503-5022-8}


\bibitem[\protect\citeauthoryear{Konishi, Maruyama, Tsubouchi, and
  Shimosaka}{Konishi et~al\mbox{.}}{2016}]%
        {Konishi:2016:CCI:2971648.2971718}
\bibfield{author}{\bibinfo{person}{Tatsuya Konishi}, \bibinfo{person}{Mikiya
  Maruyama}, \bibinfo{person}{Kota Tsubouchi}, {and} \bibinfo{person}{Masamichi
  Shimosaka}.} \bibinfo{year}{2016}\natexlab{}.
\newblock \showarticletitle{CityProphet: City-scale Irregularity Prediction
  Using Transit App Logs}. In \bibinfo{booktitle}{\emph{Proceedings of the 2016
  ACM International Joint Conference on Pervasive and Ubiquitous Computing}}
  (Heidelberg, Germany) \emph{(\bibinfo{series}{UbiComp '16})}.
  \bibinfo{publisher}{ACM}, \bibinfo{address}{New York, NY, USA},
  \bibinfo{pages}{752--757}.
\newblock
\showISBNx{978-1-4503-4461-6}


\bibitem[\protect\citeauthoryear{Lin, Feng, Lu, Li, and Jin}{Lin
  et~al\mbox{.}}{2019}]%
        {lin2019deepstn+}
\bibfield{author}{\bibinfo{person}{Ziqian Lin}, \bibinfo{person}{Jie Feng},
  \bibinfo{person}{Ziyang Lu}, \bibinfo{person}{Yong Li}, {and}
  \bibinfo{person}{Depeng Jin}.} \bibinfo{year}{2019}\natexlab{}.
\newblock \showarticletitle{DeepSTN+: Context-aware Spatial-Temporal Neural
  Network for Crowd Flow Prediction in Metropolis}. In
  \bibinfo{booktitle}{\emph{Proceedings of the AAAI Conference on Artificial
  Intelligence}}, Vol.~\bibinfo{volume}{33}. \bibinfo{pages}{1020--1027}.
\newblock


\bibitem[\protect\citeauthoryear{Lv, Duan, Kang, Li, and Wang}{Lv
  et~al\mbox{.}}{2014}]%
        {lv2014traffic}
\bibfield{author}{\bibinfo{person}{Yisheng Lv}, \bibinfo{person}{Yanjie Duan},
  \bibinfo{person}{Wenwen Kang}, \bibinfo{person}{Zhengxi Li}, {and}
  \bibinfo{person}{Fei-Yue Wang}.} \bibinfo{year}{2014}\natexlab{}.
\newblock \showarticletitle{Traffic flow prediction with big data: a deep
  learning approach}.
\newblock \bibinfo{journal}{\emph{IEEE Transactions on Intelligent
  Transportation Systems}} \bibinfo{volume}{16}, \bibinfo{number}{2}
  (\bibinfo{year}{2014}), \bibinfo{pages}{865--873}.
\newblock


\bibitem[\protect\citeauthoryear{R~Fu}{R~Fu}{2017}]%
        {RNN}
\bibfield{author}{\bibinfo{person}{L~Li R~Fu, Z~Zhang}.}
  \bibinfo{year}{2017}\natexlab{}.
\newblock \showarticletitle{Using LSTM and GRU neural network methods for
  traffic flow prediction}.
\newblock \bibinfo{journal}{\emph{Chinese Association of Automation,
  2017:324-328}} (\bibinfo{year}{2017}).
\newblock


\bibitem[\protect\citeauthoryear{Rossi, Barlacchi, Bianchini, and Lepri}{Rossi
  et~al\mbox{.}}{2019}]%
        {rossi2019modelling}
\bibfield{author}{\bibinfo{person}{Alberto Rossi}, \bibinfo{person}{Gianni
  Barlacchi}, \bibinfo{person}{Monica Bianchini}, {and} \bibinfo{person}{Bruno
  Lepri}.} \bibinfo{year}{2019}\natexlab{}.
\newblock \showarticletitle{Modelling Taxi Drivers' Behaviour for the Next
  Destination Prediction}.
\newblock \bibinfo{journal}{\emph{IEEE Transactions on Intelligent
  Transportation Systems}} (\bibinfo{year}{2019}).
\newblock


\bibitem[\protect\citeauthoryear{Sadri, Salim, Ren, Shao, Krumm, and
  Mascolo}{Sadri et~al\mbox{.}}{2018}]%
        {10.1145/3287064}
\bibfield{author}{\bibinfo{person}{Amin Sadri}, \bibinfo{person}{Flora~D.
  Salim}, \bibinfo{person}{Yongli Ren}, \bibinfo{person}{Wei Shao},
  \bibinfo{person}{John~C. Krumm}, {and} \bibinfo{person}{Cecilia Mascolo}.}
  \bibinfo{year}{2018}\natexlab{}.
\newblock \showarticletitle{What Will You Do for the Rest of the Day? An
  Approach to Continuous Trajectory Prediction}.
\newblock \bibinfo{journal}{\emph{Proc. ACM Interact. Mob. Wearable Ubiquitous
  Technol.}} \bibinfo{volume}{2}, \bibinfo{number}{4}, Article
  \bibinfo{articleno}{186} (\bibinfo{date}{Dec.} \bibinfo{year}{2018}),
  \bibinfo{numpages}{26}~pages.
\newblock


\bibitem[\protect\citeauthoryear{Seto, Sekimoto, Asahi, and Endo}{Seto
  et~al\mbox{.}}{2020}]%
        {seto2020constructing}
\bibfield{author}{\bibinfo{person}{Toshikazu Seto}, \bibinfo{person}{Yoshihide
  Sekimoto}, \bibinfo{person}{Kosuke Asahi}, {and} \bibinfo{person}{Takahiro
  Endo}.} \bibinfo{year}{2020}\natexlab{}.
\newblock \showarticletitle{Constructing a digital city on a web-3D platform:
  simultaneous and consistent generation of metadata and tile data from a
  multi-source raw dataset}. In \bibinfo{booktitle}{\emph{Proceedings of the
  3rd ACM SIGSPATIAL International Workshop on Advances in Resilient and
  Intelligent Cities}}. \bibinfo{pages}{1--9}.
\newblock


\bibitem[\protect\citeauthoryear{Song, Kanasugi, and Shibasaki}{Song
  et~al\mbox{.}}{2016}]%
        {song2016deeptransport}
\bibfield{author}{\bibinfo{person}{Xuan Song}, \bibinfo{person}{Hiroshi
  Kanasugi}, {and} \bibinfo{person}{Ryosuke Shibasaki}.}
  \bibinfo{year}{2016}\natexlab{}.
\newblock \showarticletitle{Deeptransport: Prediction and simulation of human
  mobility and transportation mode at a citywide level}. IJCAI.
\newblock


\bibitem[\protect\citeauthoryear{Song, Shao, Zhao, Cui, Shibasaki, and
  Zha}{Song et~al\mbox{.}}{2010}]%
        {song2010online}
\bibfield{author}{\bibinfo{person}{Xuan Song}, \bibinfo{person}{Xiaowei Shao},
  \bibinfo{person}{Huijing Zhao}, \bibinfo{person}{Jinshi Cui},
  \bibinfo{person}{Ryosuke Shibasaki}, {and} \bibinfo{person}{Hongbin Zha}.}
  \bibinfo{year}{2010}\natexlab{}.
\newblock \showarticletitle{An online approach:
  Learning-semantic-scene-by-tracking and tracking-by-learning-semantic-scene}.
  In \bibinfo{booktitle}{\emph{Computer Vision and Pattern Recognition (CVPR),
  2010 IEEE Conference on}}. IEEE, \bibinfo{pages}{739--746}.
\newblock


\bibitem[\protect\citeauthoryear{Tao, Zhang, Liu, and Nee}{Tao
  et~al\mbox{.}}{2018}]%
        {tao2018digital}
\bibfield{author}{\bibinfo{person}{Fei Tao}, \bibinfo{person}{He Zhang},
  \bibinfo{person}{Ang Liu}, {and} \bibinfo{person}{Andrew~YC Nee}.}
  \bibinfo{year}{2018}\natexlab{}.
\newblock \showarticletitle{Digital twin in industry: State-of-the-art}.
\newblock \bibinfo{journal}{\emph{IEEE Transactions on industrial informatics}}
  \bibinfo{volume}{15}, \bibinfo{number}{4} (\bibinfo{year}{2018}),
  \bibinfo{pages}{2405--2415}.
\newblock


\bibitem[\protect\citeauthoryear{Taylor and Letham}{Taylor and Letham}{2017}]%
        {DBLP:journals/peerjpre/TaylorL17}
\bibfield{author}{\bibinfo{person}{Sean~J. Taylor} {and}
  \bibinfo{person}{Benjamin Letham}.} \bibinfo{year}{2017}\natexlab{}.
\newblock \showarticletitle{Forecasting at Scale}.
\newblock \bibinfo{journal}{\emph{PeerJ PrePrints}}  \bibinfo{volume}{5}
  (\bibinfo{year}{2017}), \bibinfo{pages}{e3190}.
\newblock


\bibitem[\protect\citeauthoryear{Vemula, Muelling, and Oh}{Vemula
  et~al\mbox{.}}{2018}]%
        {vemula2018social}
\bibfield{author}{\bibinfo{person}{Anirudh Vemula}, \bibinfo{person}{Katharina
  Muelling}, {and} \bibinfo{person}{Jean Oh}.} \bibinfo{year}{2018}\natexlab{}.
\newblock \showarticletitle{Social attention: Modeling attention in human
  crowds}. In \bibinfo{booktitle}{\emph{2018 IEEE international Conference on
  Robotics and Automation (ICRA)}}. IEEE, \bibinfo{pages}{1--7}.
\newblock


\bibitem[\protect\citeauthoryear{Wang, Yuan, Lian, Xu, Xie, Chen, and Rui}{Wang
  et~al\mbox{.}}{2015}]%
        {10.1145/2783258.2783350}
\bibfield{author}{\bibinfo{person}{Yingzi Wang}, \bibinfo{person}{Nicholas~Jing
  Yuan}, \bibinfo{person}{Defu Lian}, \bibinfo{person}{Linli Xu},
  \bibinfo{person}{Xing Xie}, \bibinfo{person}{Enhong Chen}, {and}
  \bibinfo{person}{Yong Rui}.} \bibinfo{year}{2015}\natexlab{}.
\newblock \showarticletitle{Regularity and Conformity: Location Prediction
  Using Heterogeneous Mobility Data}. In \bibinfo{booktitle}{\emph{Proceedings
  of the 21th ACM SIGKDD International Conference on Knowledge Discovery and
  Data Mining}} (Sydney, NSW, Australia) \emph{(\bibinfo{series}{KDD ’15})}.
  \bibinfo{publisher}{Association for Computing Machinery},
  \bibinfo{address}{New York, NY, USA}, \bibinfo{pages}{1275–1284}.
\newblock
\showISBNx{9781450336642}


\bibitem[\protect\citeauthoryear{Wang, Gupta, Han, Wang, Ganlath, Ammar, and
  Tiwari}{Wang et~al\mbox{.}}{2022}]%
        {wang2022mobility}
\bibfield{author}{\bibinfo{person}{Ziran Wang}, \bibinfo{person}{Rohit Gupta},
  \bibinfo{person}{Kyungtae Han}, \bibinfo{person}{Haoxin Wang},
  \bibinfo{person}{Akila Ganlath}, \bibinfo{person}{Nejib Ammar}, {and}
  \bibinfo{person}{Prashant Tiwari}.} \bibinfo{year}{2022}\natexlab{}.
\newblock \showarticletitle{Mobility Digital Twin: Concept, Architecture, Case
  Study, and Future Challenges}.
\newblock \bibinfo{journal}{\emph{IEEE Internet of Things Journal}}
  (\bibinfo{year}{2022}).
\newblock


\bibitem[\protect\citeauthoryear{Yang, Fankhauser, Rosso, and
  Cudre-Mauroux}{Yang et~al\mbox{.}}{2020}]%
        {yang2020location}
\bibfield{author}{\bibinfo{person}{Dingqi Yang}, \bibinfo{person}{Benjamin
  Fankhauser}, \bibinfo{person}{Paolo Rosso}, {and} \bibinfo{person}{Philippe
  Cudre-Mauroux}.} \bibinfo{year}{2020}\natexlab{}.
\newblock \showarticletitle{Location Prediction over Sparse User Mobility
  Traces Using RNNs: Flashback in Hidden States!}. In
  \bibinfo{booktitle}{\emph{Proceedings of the Twenty-Ninth International Joint
  Conference on Artificial Intelligence, IJCAI-20}}.
  \bibinfo{pages}{2184--2190}.
\newblock


\bibitem[\protect\citeauthoryear{Yao, Wu, Ke, Tang, Jia, Lu, Gong, Ye, and
  Zhenhui}{Yao et~al\mbox{.}}{2018}]%
        {yao2018deep}
\bibfield{author}{\bibinfo{person}{Huaxiu Yao}, \bibinfo{person}{Fei Wu},
  \bibinfo{person}{Jintao Ke}, \bibinfo{person}{Xianfeng Tang},
  \bibinfo{person}{Yitian Jia}, \bibinfo{person}{Siyu Lu},
  \bibinfo{person}{Pinghua Gong}, \bibinfo{person}{Jieping Ye}, {and}
  \bibinfo{person}{Li Zhenhui}.} \bibinfo{year}{2018}\natexlab{}.
\newblock \showarticletitle{Deep Multi-View Spatial-Temporal Network for Taxi
  Demand Prediction}. In \bibinfo{booktitle}{\emph{The Thirty-Second AAAI
  Conference on Artificial Intelligence}}.
\newblock


\bibitem[\protect\citeauthoryear{Zhang, Zheng, and Qi}{Zhang
  et~al\mbox{.}}{2017}]%
        {zhang2017deep}
\bibfield{author}{\bibinfo{person}{Junbo Zhang}, \bibinfo{person}{Yu Zheng},
  {and} \bibinfo{person}{Dekang Qi}.} \bibinfo{year}{2017}\natexlab{}.
\newblock \showarticletitle{Deep spatio-temporal residual networks for citywide
  crowd flows prediction}. In \bibinfo{booktitle}{\emph{Thirty-First AAAI
  Conference on Artificial Intelligence}}.
\newblock


\bibitem[\protect\citeauthoryear{Zhao, Xu, Zhou, Zhao, Liu, and Zhu}{Zhao
  et~al\mbox{.}}{2018}]%
        {zhao2018prediction}
\bibfield{author}{\bibinfo{person}{Jing Zhao}, \bibinfo{person}{Jiajie Xu},
  \bibinfo{person}{Rui Zhou}, \bibinfo{person}{Pengpeng Zhao},
  \bibinfo{person}{Chengfei Liu}, {and} \bibinfo{person}{Feng Zhu}.}
  \bibinfo{year}{2018}\natexlab{}.
\newblock \showarticletitle{On prediction of user destination by sub-trajectory
  understanding: A deep learning based approach}. In
  \bibinfo{booktitle}{\emph{Proceedings of the 27th ACM International
  Conference on Information and Knowledge Management}}.
  \bibinfo{pages}{1413--1422}.
\newblock


\bibitem[\protect\citeauthoryear{Zheng, Li, Chen, Xie, and Ma}{Zheng
  et~al\mbox{.}}{2008}]%
        {Zheng:2008:UMB:1409635.1409677}
\bibfield{author}{\bibinfo{person}{Yu Zheng}, \bibinfo{person}{Quannan Li},
  \bibinfo{person}{Yukun Chen}, \bibinfo{person}{Xing Xie}, {and}
  \bibinfo{person}{Wei-Ying Ma}.} \bibinfo{year}{2008}\natexlab{}.
\newblock \showarticletitle{Understanding Mobility Based on GPS Data}. In
  \bibinfo{booktitle}{\emph{Proceedings of the 10th International Conference on
  Ubiquitous Computing}} (Seoul, Korea) \emph{(\bibinfo{series}{UbiComp '08})}.
  \bibinfo{publisher}{ACM}, \bibinfo{address}{New York, NY, USA},
  \bibinfo{pages}{312--321}.
\newblock
\showISBNx{978-1-60558-136-1}


\end{thebibliography}

\newpage
\appendix

\begin{table*}[!th]
\caption{Evaluation on Cluster-level prediction on DiDi Chengdu Data}
\label{tab:didi_experiments}
\begin{tabular}{c|cccccc}
              & GRU    & Conditional & Ensemble & Context (Max) & Context (mean) & Ours            \\ \hline
Cross Entropy & 0.5619 & 0.5630      & 0.5830       & 0.5459        & 0.5450         & \textbf{0.5447}
\end{tabular}
\end{table*}

\section{Appendix}
In this appendix, we give more details on the implementation details of our system, and another experiment on DiDi Chengdu (a publicly available dataset) for reproducibility.

\subsection{Implementation Details}
\begin{figure}[!h]
\centering
\includegraphics[width=\linewidth]{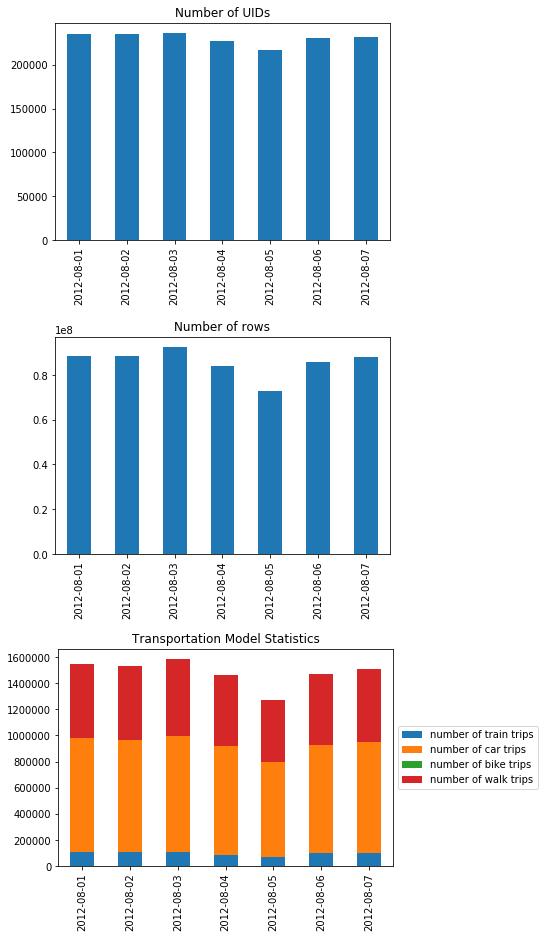}
\caption{Illustration of the number of user IDs (left), number of rows (middle) and the distribution of the number of trips with respect to different transportation trips in our fine-grained node-level trajectory database.}
\label{fig:basic_statistics}
\end{figure}
\subsubsection{Experimental Setups}
In our previous experiments, we set the time interval to be 5 min for trajectory representation, and used the most recent 1h (12 steps) trajectories to make a 1h-ahead prediction every 15 minutes. The cluster and time embedding dimensions were set to 128 and 64 respectively. Moreover, a two-layered GRU with a hidden size of 256 was utilized to model the sequential pattern for individual trajectories and a single-layered GRU with a hidden size of 64 was utilized to model the sequential pattern of the crowd context. A Multi-class MLP layer was implemented as a two-layered network, with a 256 dimension latent layer.

\subsubsection{Data Details}
As shown in Fig. \ref{fig:basic_statistics}, we show one-week number of unique user IDs, number of rows (every time our node-level trajectories passing a transportation node, we write one row describing the timestamp and the node in the database), and the distribution of the trips with respect to different transportation model.

We determined the home location each user in the dataset, which was compared with census data on 1km grid sections, and estimated the linear relationship as: $$N_{GPS} \,=\, 0.0063\, *\, N_{census}\, +\, 0.74, \: R^2\, =\, 0.79$$ where $N_{GPS}$ is the population estimated from GPS dataset, $N_{census}$ is the population given by the national census data, and $R^2$ is the coefficient of determination.

\subsection{Additional Experiment on DiDi Chengdu (public dataset)}
Because the dataset used in this paper cannot be published due to privacy concerns, to help researchers to reproduce this paper, we conduct additional experiment on DiDi Chengdu City Second Ring Road Regional Trajectory Data Set in Oct and Nov 2016 \footnote{\url{https://gaia.didichuxing.com}}. We have open-sourced the experiment scripts with detailed descriptions and all baseline models (in the cluster-level prediction) we used in this paper \footnote{\url{https://github.com/fanzipei/crowd-context-prediction/tree/master}}. The code has been modified to work with DiDi Chengdu dataset, so the researchers can reproduce our system easily.

\begin{figure}[!h]
\centering
\includegraphics[width=\linewidth]{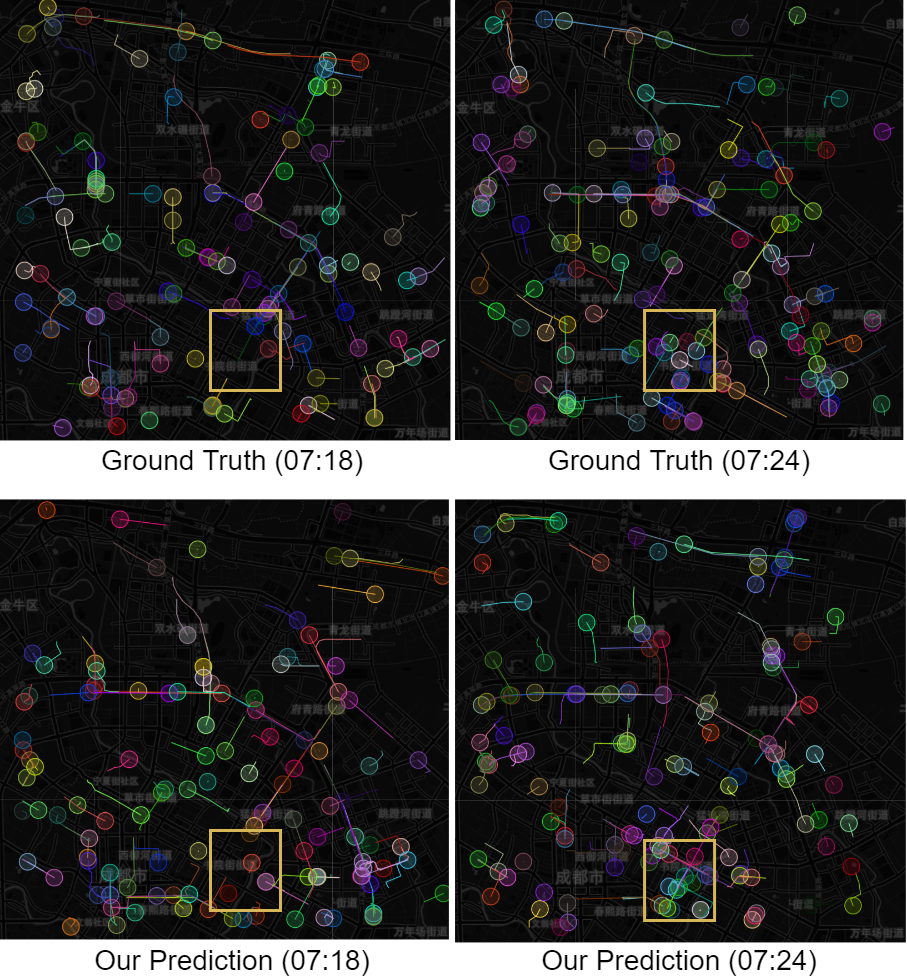}
\caption{DiDi taxi trajectory 6 min / 12 min ahead prediction. A significant increase of taxi trajectories in the yellow region can be seen from the upper row, and our predictor can successfully predict the increase of the taxi trajectories.}
\label{fig:didi_fine_grained}
\end{figure}

In this experiment, we split the DiDi Chengdu dataset into training set (from 2016.10.01 to 2016.11.14) and testing set (from 2016.11.15 to 2016.11.30). Considering the target region of this dataset is smaller and the taxi trips do not have a continuous observation of the user, we cropped a shorter period (all days from 7 a.m. to 11 a.m.) with a smaller interval (1 min). We take the order IDs as the pseudo "user IDs" use the latest 12 time steps observations of trajectories to predict their future movements (12 min ahead). Table \ref{tab:didi_experiments} shows the performance of our cluster-level prediction on DiDi dataset. Our proposed predictor outperforms all the baseline models. Same to our previous experiments, \textbf{Context (Mean)} is the second-best predictor, still performs slightly better than the \textbf{Context (Max)}. \textbf{Conditional} and \textbf{Ensemble} predictors do not achieve a good prediction results in this experiment. One probable reason is the training dataset include too many irregular patterns. Note that the whole first week in Oct is the national holidays in China, and thus we labeled them all as holidays. \textbf{Conditional}
and \textbf{Ensemble} do not have sufficient ability to distinguish these holidays from normal holidays, which leads to a worse prediction performance.

In the fine-grained prediction stage, we skipped the stay-move detection and transportation mode classification steps since all the trajectories are all moving cars. The original GPS trajectories can be directly taken as fine-grained trajectories because it has a very short time interval (3-4 seconds) and has already been snapped to the road network. Our proposed two-stage fine-grained mobility prediction algorithm can also handle this type of data without minor modifications. Figure \ref{fig:didi_fine_grained} shows our fine-grained prediction results. Notably, our two-stage fine-grained prediction can not only predict complete trajectories for each individuals at a low cost (on average it takes about 7 seconds on each prediction for over 40K trajectories), but also successfully predict the aggregated trajectory patterns, such as the gathering pattern at a office area in Chengdu as shown on the right column in Figure \ref{fig:didi_fine_grained}.
\end{document}